\documentclass[preprint2]{aastex61}
\usepackage{graphicx}
\usepackage{psfig}
\usepackage{natbib}

\newcommand{\gsim}{\mbox{\hspace{.2em}\raisebox{.5ex}{$>$}\hspace{-.8em}\raisebox{-.5ex}{$\sim$}\hspace{.2em}}}
\newcommand{\lsim}{\mbox{\hspace{.2em}\raisebox{.5ex}{$<$}\hspace{-.8em}\raisebox{-.5ex}{$\sim$}\hspace{.2em}}}

\newcommand{\E}[1]{\times 10^{#1}}
\newcommand{\twCO}{$^{12}$CO}  \newcommand{\thCO}{$^{13}$CO}

\newcommand{\HII}{\mbox{H\,\textsc{ii}}}

      \newcommand{\ps}{\,{\rm s}^{-1}}
       
    \newcommand{\km}{\,{\rm km}}

\begin{document}

\title{
Molecular Environments of Three Large Supernova Remnants in the Third
Galactic Quadrant:
G205.5$+$0.5, G206.9$+$2.3, and G213.0$-$0.6
}

\shorttitle{Molecular Environment of three Large SNRs}

\correspondingauthor{Yang Su}
\email{yangsu@pmo.ac.cn}

\author[0000-0002-0197-470X]{Yang Su}
\affil{Purple Mountain Observatory and Key Laboratory of Radio Astronomy,
Chinese Academy of Sciences, Nanjing 210008, China}

\author{Xin Zhou}
\affiliation{Purple Mountain Observatory and Key Laboratory of Radio Astronomy,
Chinese Academy of Sciences, Nanjing 210008, China}

\author{Ji Yang}
\affiliation{Purple Mountain Observatory and Key Laboratory of Radio Astronomy,
Chinese Academy of Sciences, Nanjing 210008, China}

\author{Xuepeng Chen}
\affiliation{Purple Mountain Observatory and Key Laboratory of Radio Astronomy,
Chinese Academy of Sciences, Nanjing 210008, China}

\author{Yang Chen}
\affiliation{Department of Astronomy, Nanjing University,
Nanjing 210023, China}
\affiliation{Key Laboratory of Modern Astronomy and Astrophysics,
Nanjing University, Ministry of Education, Nanjing 210023, China}

\author{Yi Liu}
\affiliation{Purple Mountain Observatory and Key Laboratory of Radio Astronomy,
Chinese Academy of Sciences, Nanjing 210008, China}

\author{Hongchi Wang}
\affiliation{Purple Mountain Observatory and Key Laboratory of Radio Astronomy,
Chinese Academy of Sciences, Nanjing 210008, China}

\author{Chong Li}
\affiliation{Purple Mountain Observatory and Key Laboratory of Radio Astronomy,
Chinese Academy of Sciences, Nanjing 210008, China}
\affiliation{University of Chinese Academy of Sciences, 
19A Yuquan Road, Shijingshan District, Beijing 100049, China}

\author{Shaobo Zhang}
\affiliation{Purple Mountain Observatory and Key Laboratory of Radio Astronomy,
Chinese Academy of Sciences, Nanjing 210008, China}

\begin{abstract}
We present CO observations toward three large supernova remnants 
(SNRs) in the third Galactic quadrant using the Purple Mountain 
Observatory Delingha 13.7 m millimeter-wavelength telescope. The observations 
are part of the high-resolution CO survey of the Galactic plane 
between Galactic longitudes $l=-10^{\circ}$ to $250^{\circ}$ 
and latitudes $b=-5^{\circ}$ to $5^{\circ}$. CO emission was 
detected toward the three SNRs: G205.5$+$0.5 (Monoceros Nebula), 
G206.9$+$2.3 (PKS 0646$+$06), and G213.0$-$0.6. Both of SNRs 
G205.5$+$0.5 and G213.0$-$0.6 exhibit the morphological agreement
(or spatial correspondences) between the remnant and the surrounding 
molecular clouds (MCs), as well as kinematic signatures of shock 
perturbation in the molecular gas. We confirm that the two 
SNRs are physically associated with their ambient MCs and the 
shock of SNRs is interacting with the dense, clumpy molecular gas. 
SNR G206.9$+$2.3, which is close to the northeastern edge of the
Monoceros Nebula, displays the spatial coincidence with 
molecular partial shell structures at $V_{\rm LSR}\sim15\km\ps$. 
While no significant line broadening has been detected within or 
near the remnant, the strong morphological correspondence between 
the SNR and the molecular cavity implies that SNR G206.9$+$2.3 is
probably associated with the CO gas and is evolving in the 
low-density environment. The physical features of individual SNRs, 
together with the relationship between SNRs and their nearby objects, 
are also discussed.
\end{abstract}


\section{INTRODUCTION}
Most core-collapse supernovae (SNe) from high-mass stars 
may explode promptly and are thought to be associated with the 
molecular clouds (MCs) from which they were born. Therefore, the 
subsequent supernova remnants (SNRs) may evolve close to their 
parental MCs or the nearby molecular environment. So far, there 
are several signposts to show the physical contact/interaction 
between SNRs and their molecular environments, e.g., the presence of
1720 MHz OH maser emission, molecular line broadenings or
asymmetric profiles, high molecular line ratios between different 
excitation states, detection of H$_2$ and/or [Fe II] lines 
in near-infrared, specific infrared colors within an SNR, and often 
used morphological correspondence in multiwavelength with SNR 
features \citep[see][]{2010ApJ...712.1147J}. 
It is also important to study the interplay of SNR--MC systems 
for investigating various physical and astrophysical processes
therein \citep[e.g.,][]{2014IAUS..296..170C,2015A&ARv..23....3D,2015SSRv..188..187S}.

There are currently 294 SNRs known in the Milky Way 
\citep{2014BASI...42...47G} and about 70 of them are 
confirmed or suggested to be associated with MCs 
\citep[refer to, e.g.,][]{2010ApJ...712.1147J,2014IAUS..296..170C}.
CO, as an indicator of H$_2$, is the most widely used tracer 
of MCs and it is very common and useful when investigating 
the nature of the molecular ISM in galaxies. CO observations 
also play a key role in studying the molecular environment and 
physical characteristics of SNR--MC interacting systems. To date, 
three large-scale, systematic CO surveys were performed to investigate the 
SNR--MC association \citep{1986ApJ...309..804H,2012Ap&SS.342..389J,
2016ApJ...816....1K}. Recently, several individual 
SNR--MC systems were also investigated using CO lines (\citealp
[e.g., SNR W44 by][]{2014A&A...569A..81A}; 
\citealp[SNR Cassiopeia A by][]{2014ApJ...796..144K};
\citealp[SNR IC 443 by][]{2014ApJ...788..122S};
\citealp[SNR G22.7$-$0.2 by][]{2014ApJ...796..122S};
\citealp[SNR G127.1$+$0.5 by][]{2014ApJ...791..109Z};
\citealp[SNR G18.8$+$0.3 by][]{2015A&A...580A..51P};
\citealp[SNR G357.7$+$0.3 by][]{2017ApJ...834...12R};
\citealp[SNR RCW 86 by][]{2016arXiv160607745S};
\citealp[SNR Tycho by][]{2016ApJ...826...34Z} 
and \citealp{2016arXiv160805329C}; and
\citealp[SNR HB 3 by][]{2016ApJ...833....4Z}).
Nevertheless, few studies have been conducted to investigate the
SNRs' molecular environment in the third Galactic quadrant.

The Milky Way Imaging Scroll Painting (MWISP) 
project\footnote{http://www.radioast.nsdc.cn/mwisp.php} is
a large \twCO\ ($J$=1--0), \thCO\ ($J$=1--0), 
and C$^{18}$O ($J$=1--0) survey of the northern Galactic plane. 
The systematic and unbiased CO survey, which has high spatial 
($\sim50''$) and velocity ($\sim0.2\km\ps$) resolution, 
will provide the detailed distributions and fully sampled images 
of molecular gas from $l=-10^{\circ}$ to $250^{\circ}$ and 
latitudes from $b=-5^{\circ}$ to $5^{\circ}$. 
The high-quality CO data also provide us with a good opportunity to study 
the molecular environment of extended SNRs from large scale to 
small scale. Especially for SNRs with large angular size, the
unbiased CO survey can offer detailed molecular gas information
on SNR--MC interactions.

In this paper, we aim to investigate the molecular environment 
of three large SNRs: G205.5$+$0.5 (Monoceros Nebula), 
G206.9$+$2.3 (PKS 0646$+$06), and G213.0$-$0.6 from our
new CO survey. 
The three SNRs are all in the third 
Galactic quadrant and their angular sizes are 
$\gsim1^{\circ}$. Benefiting from the large-scale spectroscopic 
mapping of the CO survey, we analyzed the correlation between 
these three remnants and their molecular environment mainly
from SNR--MC's morphological correspondences (e.g., molecular 
arc, shell, cavity, interface, and coincidence in multiwavelength 
emission, etc.), molecular line diagnostics (e.g., broad 
molecular line features or asymmetric line profiles),
and special kinematic features (e.g., velocity gradient revealed  
in the position$-$velocity (PV) diagram). 

In Section 2, we describe the CO observation and the data 
reduction, and in Section 3 we present our results and discussions 
for individual SNRs. Finally, a summary is given in Section 4.

\section{CO data from MWISP}
The MWISP project, which started in 2011, is a large CO
survey using the 13.7 m millimeter-wavelength telescope located
at Delingha in China. The \twCO\ ($J$=1--0), \thCO\ ($J$=1--0), 
and C$^{18}$O ($J$=1--0) observations are performed simultaneously
with the nine-beam Superconducting Spectroscopic Array Receiver 
(SSAR) system \citep{Shan}. 
A Fast Fourier Transform Spectrometer (FFTS) with a total 
bandwidth of 1~GHz provides 16,384 channels, resulting in a spectral 
resolution of 61 kHz, equivalent to a velocity resolution of 
$\sim0.17\km\ps$ at 110 GHz. The intermediate frequency (IF) band 
is 2.64$\pm$0.5 GHz so that three CO isotope lines (\twCO\ line at
115.271~GHz, \thCO\ line at 110.201~GHz, and C$^{18}$O line at
109.782~GHz) can be covered 
by the 1 GHz band when the local oscillator (LO) is at 112.6 GHz.
We should specify that the FFTS is sideband separated
and \twCO\ line is in the upper while \thCO\ and C$^{18}$O
are in the lower band.
The half-power beamwidth (HPBW) of the telescope is about
$50''$ and the pointing accuracy is better than $5''$ in all observing
epochs. The main beam efficiency and other useful parameters can be found
from the status report\footnote{http://www.radioast.nsdc.cn/mwisp.php}.

All covered sky will be divided into 10,941 cells and each cell 
with a dimension of 30$'\times30'$ is scanned along the Galactic 
longitude and the Galactic latitude 
to reduce the fluctuation of noise. Observations are conducted 
in position-switch On-The-Fly (OTF) mode, scanning each cell at
a rate of $50''$ (or $75''$) per second with a dump time of
0.3 s (or 0.2 s). Before the observation, the background 
reference region ($8'\times8'$) near the cell has been carefully
checked ($\lesssim$0.3~K in \twCO\ emission at any points) to ensure no 
or little emission from the off region. 
In addition, the water vapor in the terrestrial atmosphere 
contributes noise in our CO spectra. Our survey 
requires relatively dry and stable atmosphere conditions to guarantee the 
quality of the data. As a result, typical system temperatures 
are within $\sim$190--350 K at the upper sideband and $\sim$140--230 K 
at the lower sideband depending on weather conditions, cell's elevations,
and off positions.

All data were reduced using the GILDAS/CLASS
package\footnote{http://www.iram.fr/IRAMFR/GILDAS}.
After the first order (linear) baseline fitting
and mosaicing the image, the final three-dimensional (3D) cube data 
were obtained with a grid spacing of $30''$. The typical 
sensitivity (RMS) is about 0.5 K for \twCO\ ($J$=1--0) at the 
channel width of 0.16$\km\ps$ and 0.3 K for \thCO\ ($J$=1--0) 
and C$^{18}$O ($J$=1--0) at 0.17$\km\ps$. As of this writing 
the MWISP project has completed about half of its planned area 
of coverage and the three large SNRs G205.5$+$0.5, G206.9$+$2.3, 
and G213.0$-$0.6, including their nearby regions,  were 
all mapped from 2012 January to 2016 April.
Generally, C$^{18}$O ($J$=1--0) 
emission is too weak to give any further details 
toward these SNRs. Our discussions will focus primarily 
on \twCO\ and \thCO\ data.

\section{RESULTS AND DISCUSSIONS}
\subsection{SNR G205.5$+$0.5 (Monoceros Nebula)} G205.5$+$0.5 
(Monoceros Nebula) is a very large SNR with a size of $\sim220'$. 
\cite{2012A&A...545A..86X} showed that the radio spectral
index ($\alpha$) of the remnant is about $-0.41$, derived from 21 cm, 
11 cm, and 6 cm radio continuum data, which is generally consistent 
with the result of previous works 
\citep{1975AJ.....80..437D,1982A&A...109..145G}. Throughout this 
paper, all directions are described in Galactic coordinates to 
match the MWISP CO survey. The bright Rosette Nebula (Sh 2-275) 
and open cluster NGC 2264 (Sh 2-273) are located near the southern 
and northwestern boundary of the remnant, respectively. 
\cite{1986ApJ...301..813O} suggested that the SNR is probably 
located behind the Rosette Nebula because of the absorption of the 
remnant's nonthermal emission by the southern \HII\ region Sh 2-275.
\cite{2012A&A...545A..86X} also suggested that the SNR is probably 
associated with the Rosette Nebula because of the identified western 
partial shell structures at $\sim15\km\ps$ using the new Arecibo 
HI 21 cm data. The remnant displays contrasting optical 
filamentary structures mainly on its northwestern, western, and 
southwestern edges. These sharp optical filaments, especially toward 
the western boundary of the remnant, are well correlated with the 
bright radio shells \citep{1978A&AS...31..271D} and the X-ray 
emission regions \citep{1985MNRAS.213P..15L,1986MNRAS.220..501L}. 
According to the Sino-German 6 cm survey, \cite{2011A&A...529A.159G}
found that the magnetic field is mainly aligned with the western 
shell of the remnant.

\begin{figure*}
\includegraphics[trim=5mm 0mm 0mm 100mm,scale=0.87,angle=0]{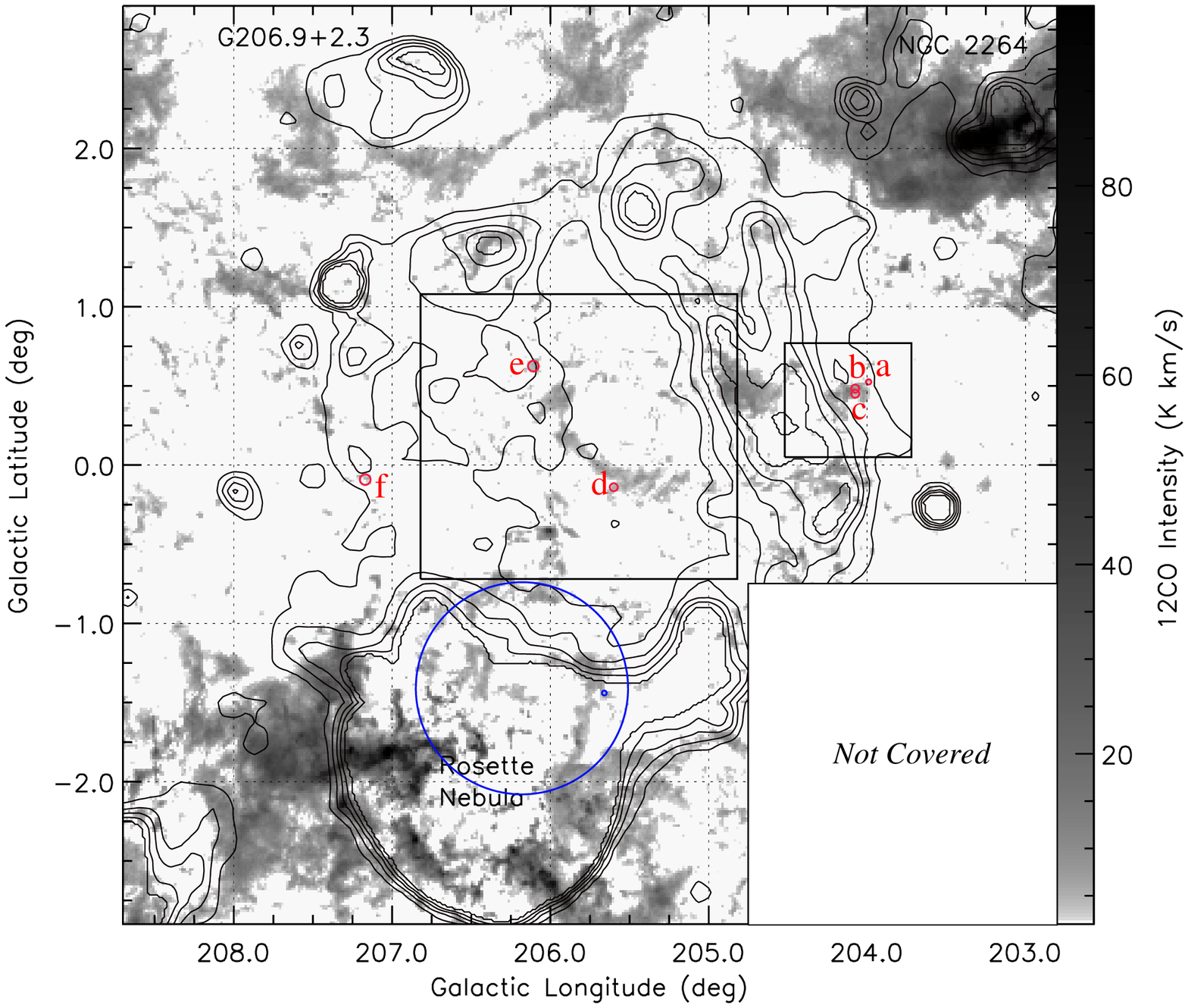}
\caption{
Integrated \twCO\ ($J$=1--0) emission toward SNR G205.5$+$0.5 
from $-10$ to 50~km~s$^{-1}$, overlaid with radio continuum contours 
from the Effelsberg 21 cm survey \citep{1997A&AS..126..413R}.
The radio contour levels are at 340, 480, 620, 760, and 900 mK.
The two rectangles indicate regions shown in Figures \ref{Moncen} and 
\ref{Monsub}, respectively. Six red circles, which are denoted 
with letters a--f, indicate positions of shocked gas (see spectra 
in Figure \ref{Monspec}). Sizes of these red circles are enlarged to two
times with respect to their angular sizes (1$'$--2$'$). The nearby sources 
of NGC 2264 and Rosette Nebula, as well as SNR G206.9$+$2.3, are also labeled.
The little and large blue circles indicate the point TeV source, 
HESS J0632$+$057 \citep{2007A&A...469L...1A}, and extended or confused source,
3EG J0634$+$0521 \citep{1999ApJS..123...79H,2008A&A...489..849C},
respectively.
\label{Mon12}}
\end{figure*}

In our large CO survey with spatial resolution better than 
$\sim1'$, we detected molecular gas over a wide velocity range 
from $-10$ to 50$\km\ps$ toward SNR G205.5$+$0.5 (Figure \ref{Mon12}). 
Obviously, two large giant molecular cloud (GMC) complexes are 
associated with the southern Rosette Nebula (MCs in the 
lower-left corner of Figure \ref{Mon12}, 
$V_{\rm LSR}\sim$3--18$\km\ps$) and the northwestern NGC 2264 (MCs 
in the upper-right corner of Figure \ref{Mon12}, 
$V_{\rm LSR}\sim$3--12$\km\ps$), respectively. Our unbiased 
CO survey revealed huge amounts of molecular gas in the 
velocity range from 25 to 50$\km\ps$. The 25--50$\km\ps$ MCs, 
which have small sizes and weak CO emission, are distributed 
widely and separately over the field of view (FOV). The CO 
emission of the 25--50$\km\ps$ components along the line of sight 
(LOS) is probably from the distant molecular gas (e.g., the Perseus 
and/or the Outer arm in the direction). We did not find any evidence 
of physical connection between the $V_{\rm LSR}\gsim$25$\km\ps$ 
molecular gas and the remnant. We will not discuss the 
MC components and we will mainly focus on the 
$V_{\rm LSR}\lsim$25$\km\ps$ MCs in this section.

When the SNR's shock is encountering a dense medium,
the shock will compress, heat, accelerate, and even dissociate 
molecules, which will lead to a wide variety of observable effects. 
The material may carry enough momentum to
accelerate the molecular gas to broad velocity width.
Generally, the asymmetric line profiles from \twCO\ emission
usually imply the turbulent molecular gas because such emission 
is more easily influenced by the shock. On the contrary, 
\thCO\ emission is optically thin, which is not heavily influenced 
by SNR's shock. 
In another words, \twCO\ emission is from the enveloping 
layer of low-density gas and is readily perturbed by local turbulence
while \thCO\ emission is not so.
The line peak of \thCO\ emission is thus a 
good indicator of the LSR velocity of unperturbed clouds. 
As a result, the asymmetric line profiles of \twCO\ can be 
picked out from the LSR velocity range of the \thCO\ component
(e.g., refer to the blue broadening from Figures \ref{Monpv} 
and \ref{Monspec} for shocked gas a; the red broadening from Figures 
\ref{Monpvde} and \ref{Monspec} for shocked gas d--e).

\begin{figure*}
\includegraphics[trim=-70mm 0mm 0mm 120mm,scale=0.5,angle=0]{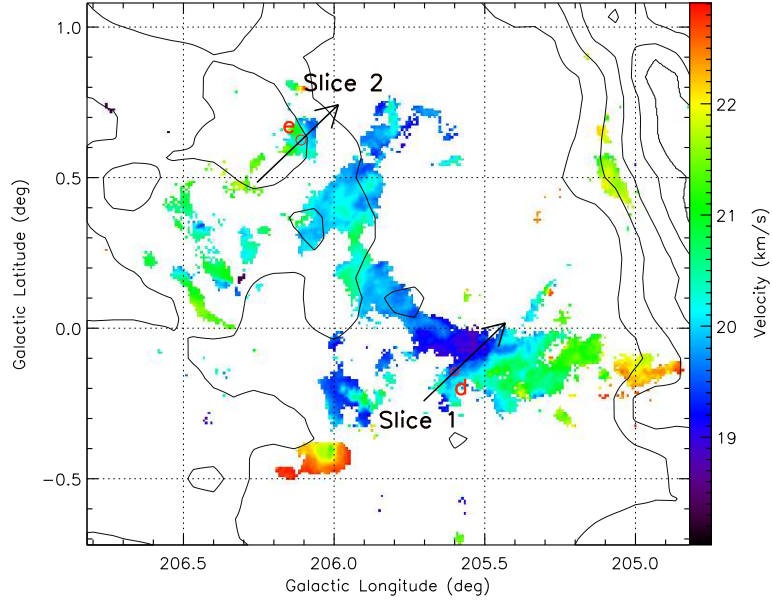}
\caption{
Intensity-weighted \twCO\ ($J$=1--0) mean velocity (first moment) map of
MCs toward the center of SNR G205.5$+$0.5 in the interval of 
18--23~km~s$^{-1}$, overlaid with the same radio contours as in 
Figure \ref{Mon12}. 
The two arrows of Slice 1 and Slice 2 indicate the PV 
slice shown in Figure \ref{Monpvde}.
Positions and sizes of shocked gas (see spectra in 
Figure \ref{Monspec}) are also labeled.
\label{Moncen}}
\end{figure*}

\begin{figure*}
\centerline{
\includegraphics[trim=-5mm 0mm 0mm 0mm,scale=0.3,angle=0]{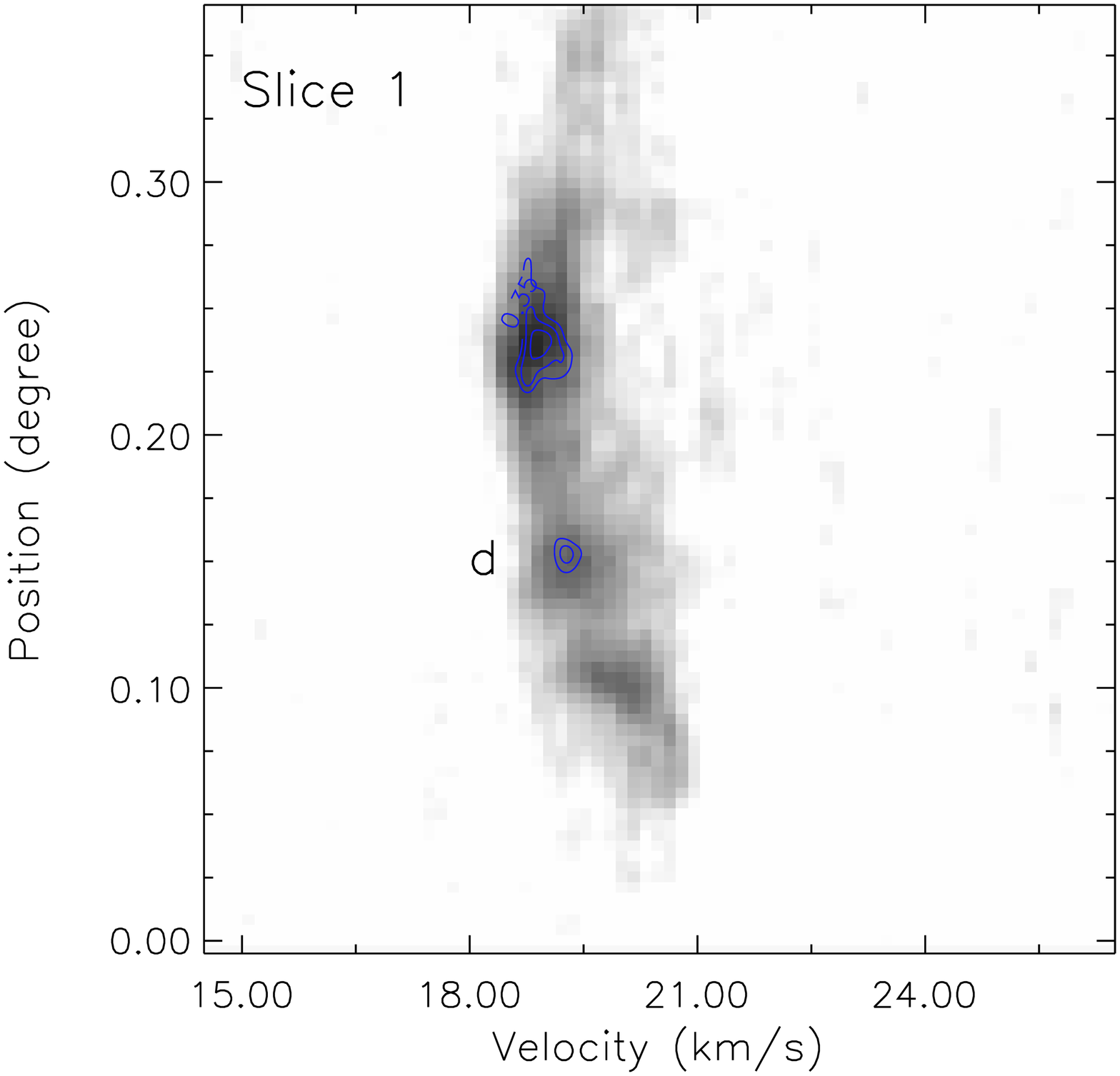}
\includegraphics[trim=-5mm 0mm 0mm 0mm,scale=0.3,angle=0]{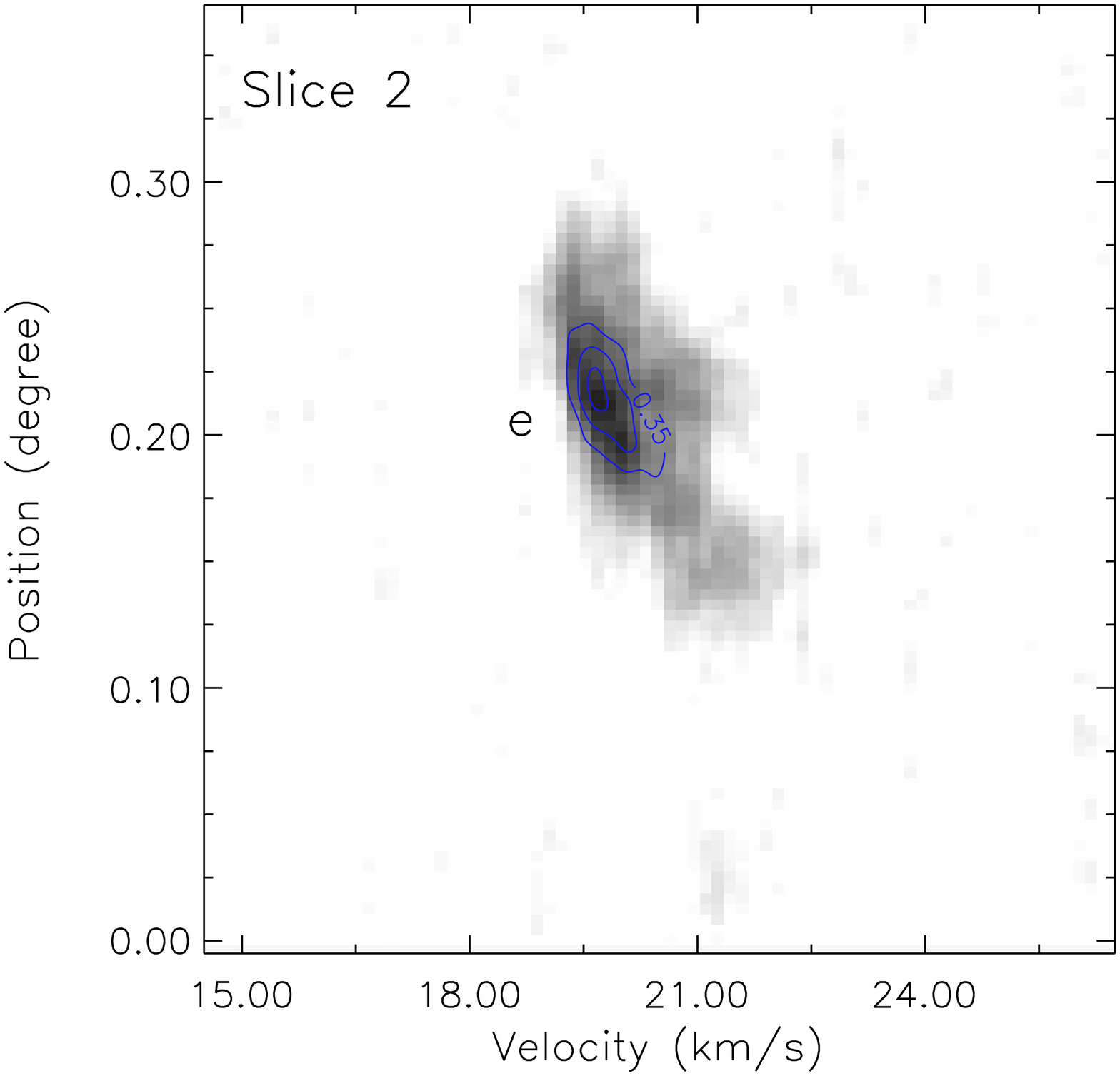}}
\caption{
Left panel: PV diagram of \twCO\ ($J$=1--0) emission along slice 1,
overlaid with the blue contours of \thCO\ emission. The overlaid contours
levels are at 0.35 (3$\sigma$), 0.47, and 0.58 K.
The PV slice has a length of 22\farcm5
(from ($l=$205\fdg703, $b=-$0\fdg242) to ($l=$205\fdg432, $b=$0\fdg017))
and a width of 2\farcm5 (see the arrow slice 1 in Figure \ref{Moncen}).
Right panel: PV diagram of \twCO\ ($J$=1--0) emission along slice 1,
overlaid with the blue contours of \thCO\ emission. The overlaid contours
levels are at 0.35 (3$\sigma$), 0.70, and 1.05 K.
The PV slice has a length of 22\farcm5
(from ($l=$206\fdg256, $b=$0\fdg484) to ($l=$205\fdg986, $b=$0\fdg742))
and a width of 2\farcm5 (see the arrow slice 2 in Figure \ref{Moncen}).
\label{Monpvde}}
\end{figure*}

Actually, several signatures of SNR--MC interaction toward the 
remnant are revealed in our CO observations. 
First, the molecular gas at $V_{\rm LSR}\sim$19$\km\ps$ (in the 
interval of 18--23$\km\ps$) seems to be concentrated in the 
center region of SNR G205.5$+$0.5 (Figure \ref{Moncen}) and this 
molecular gas exhibits weak redshifted broadening in \twCO\ emission 
(e.g., the typical spectra of d and e in Figure \ref{Monspec}). 
The broadening 
of the two positions is not very large ($\sim$2--3 $\km\ps$ from 
the \thCO\ peak) but can be easily discerned from the asymmetric 
\twCO\ profile. 
It is worth noting that the weak emission in \twCO\ is systematic 
redshifted with respect to the unperturbed clouds traced by \thCO\
(Figure \ref{Monpvde}), which indicates the \twCO\ wing component in the
velocity interval of $\sim$19--22$\km\ps$ and the
\thCO\ component at a slightly low LSR velocity of $\sim$19$\km\ps$ 
are apparently connected. Therefore, the weak emission seen in
the redshifted \twCO\ wing (d--f in Figure \ref{Monspec}) and the \thCO\
emission is from the coherent molecular gas, but not from 
the separated source with slightly different distances along the LOS.
Generally, the redshifted broadening of d--f is not very large.
High J CO lines (e.g., 2--1 and/or 3--2) may be useful for investigating 
these regions in the future.

\begin{figure*}
\includegraphics[trim=-15mm 0mm 0mm 165mm,scale=0.75,angle=0]{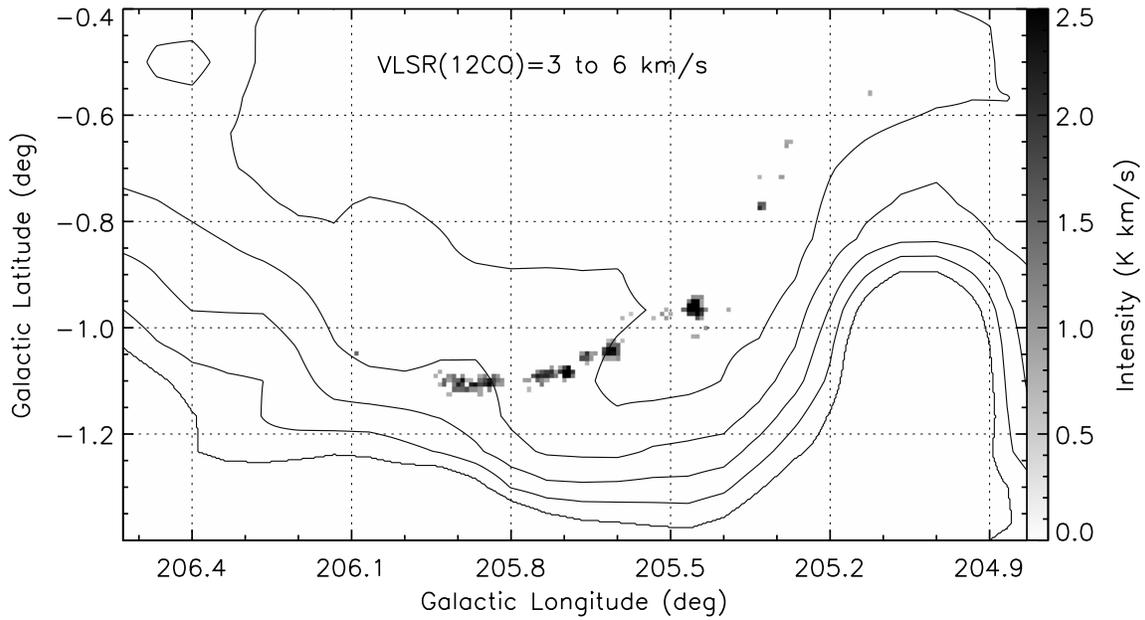}
\caption{
Integrated \twCO\ ($J$=1--0) emission toward the southwestern boundary of 
SNR G205.5$+$0.5 from 3 to 6~km~s$^{-1}$,
overlaid with the same radio contours as in Figure \ref{Mon12}.
\label{Monsm36}}
\end{figure*}

\begin{figure*}
\includegraphics[trim=-15mm 0mm 0mm 165mm,scale=0.75,angle=0]{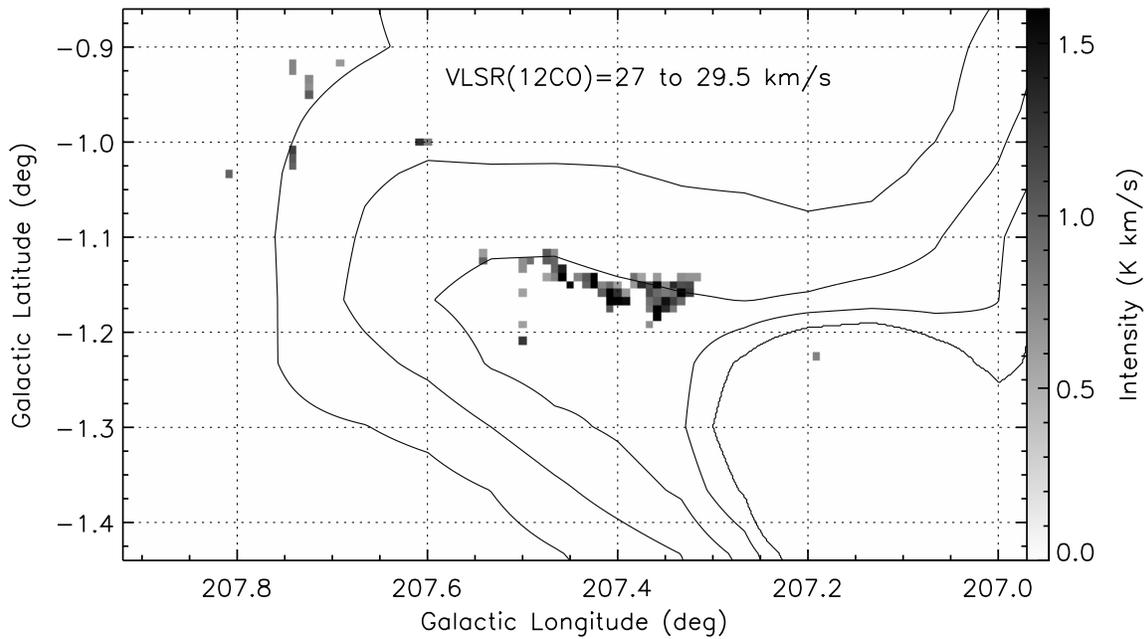}
\caption{
Integrated \twCO\ ($J$=1--0) emission toward the southeastern shell of
SNR G205.5$+$0.5 from 27 to 29.5~km~s$^{-1}$,
overlaid with the same radio contours as in Figure \ref{Mon12}.
\label{Monsm27295}}
\end{figure*}

Second, two partial shell structures traced by \twCO\ 
in $V_{\rm LSR}$=3--6$\km\ps$ and 
$V_{\rm LSR}$=27--29.5$\km\ps$ seem to be correlated with the 
southwestern (Figure \ref{Monsm36}) and the southeastern 
(Figure \ref{Monsm27295}) radio emission, respectively.
Especially, the partial shell structure revealed by 
the 3--6$\km\ps$ molecular component is as long as 52$'$, or
$\sim$24 pc at a distance of 1.6 kpc (see below).
We suggest that the molecular structures are probably related 
to the remnant; nevertheless, the CO emission of the two shells 
is too weak to be studied in detail.

\begin{figure*}
\centerline{
\includegraphics[trim=-5mm 0mm 0mm 120mm,scale=0.43,angle=0]{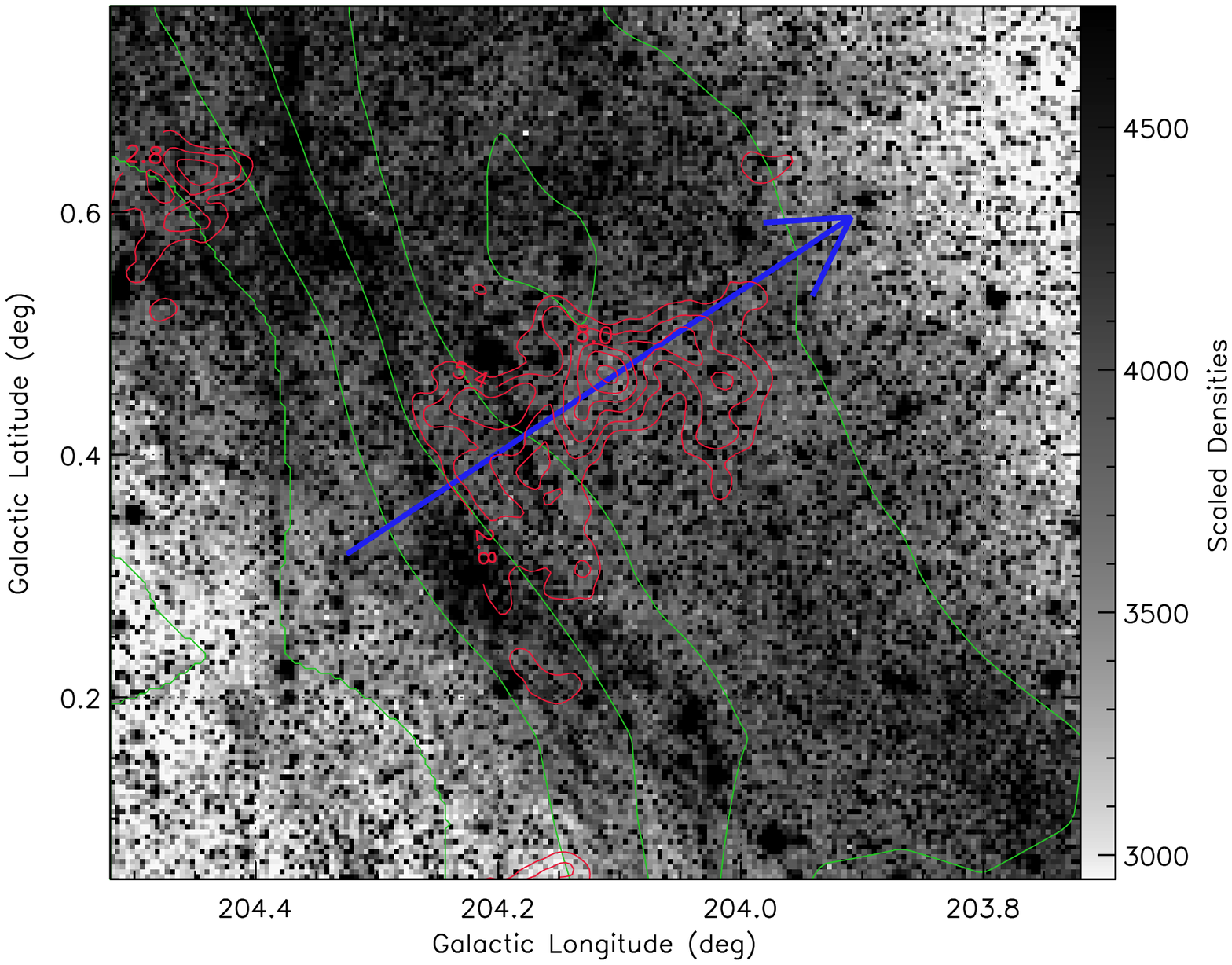}
\includegraphics[trim=-5mm 0mm 0mm 120mm,scale=0.43,angle=0]{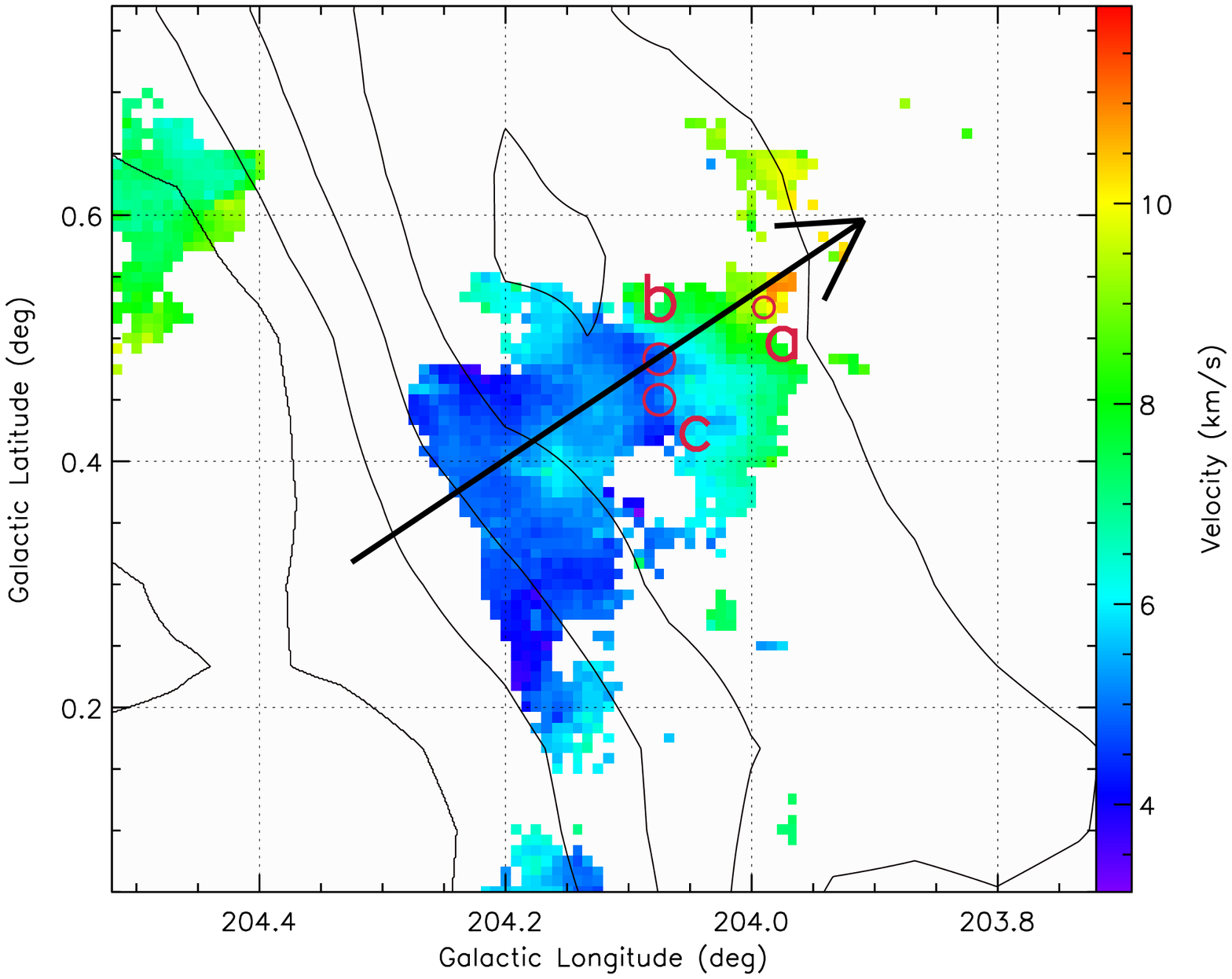}}
\caption{
Left panel: optical image \citep[the Digitized Sky Survey,][]{1990AJ.....99.2019L}
toward the western boundary of SNR G205.5$+$0.5,
overlaid with green radio contours as in Figure \ref{Mon12}. 
Red contours show the \twCO\ ($J$=1--0) distribution 
(in units of K~$\km\ps$) integrated from 3 to 12~km~s$^{-1}$.
The arrow indicates the PV slice shown in Figure \ref{Monpv}.
Right panel: intensity-weighted \twCO\ ($J$=1--0) mean velocity 
(first moment) map of MCs in the interval of 3--12~km~s$^{-1}$, 
overlaid with the same radio contours as in the left panel. 
Positions and sizes of shocked gas (see spectra in Figure \ref{Monspec}) 
are also labeled.
\label{Monsub}}
\end{figure*}

\begin{figure*}
\includegraphics[trim=-90mm 0mm 0mm 20mm,scale=0.4,angle=0]{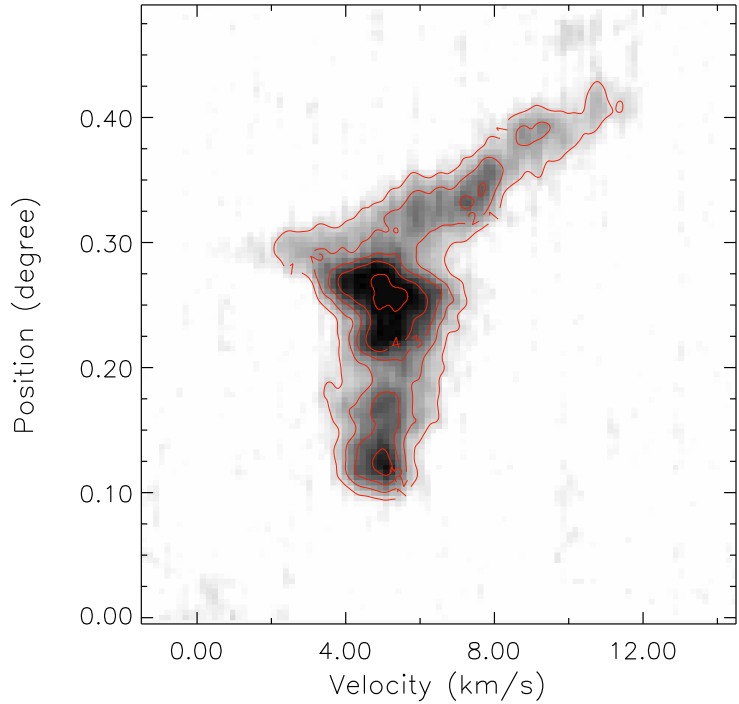}
\caption{
PV diagram of \twCO\ ($J$=1--0) emission in the western radio boundary
of SNR G205.5$+$0.5 (see the right rectangle in Figure \ref{Mon12}). 
The PV slice with a length of 30$'$ 
(from ($l=$204\fdg325, $b=$0\fdg318) to ($l=$203\fdg909, $b=$0\fdg596))
and a width of 2\farcm5 is just perpendicular 
to the SNR's radio shell (see the arrow in Figure \ref{Monsub}).
The red contours indicate the emission of \twCO\ in units of K.
\label{Monpv}}
\end{figure*}

\begin{figure*}
\includegraphics[trim=0mm 0mm 0mm -10mm,scale=0.32,angle=0]{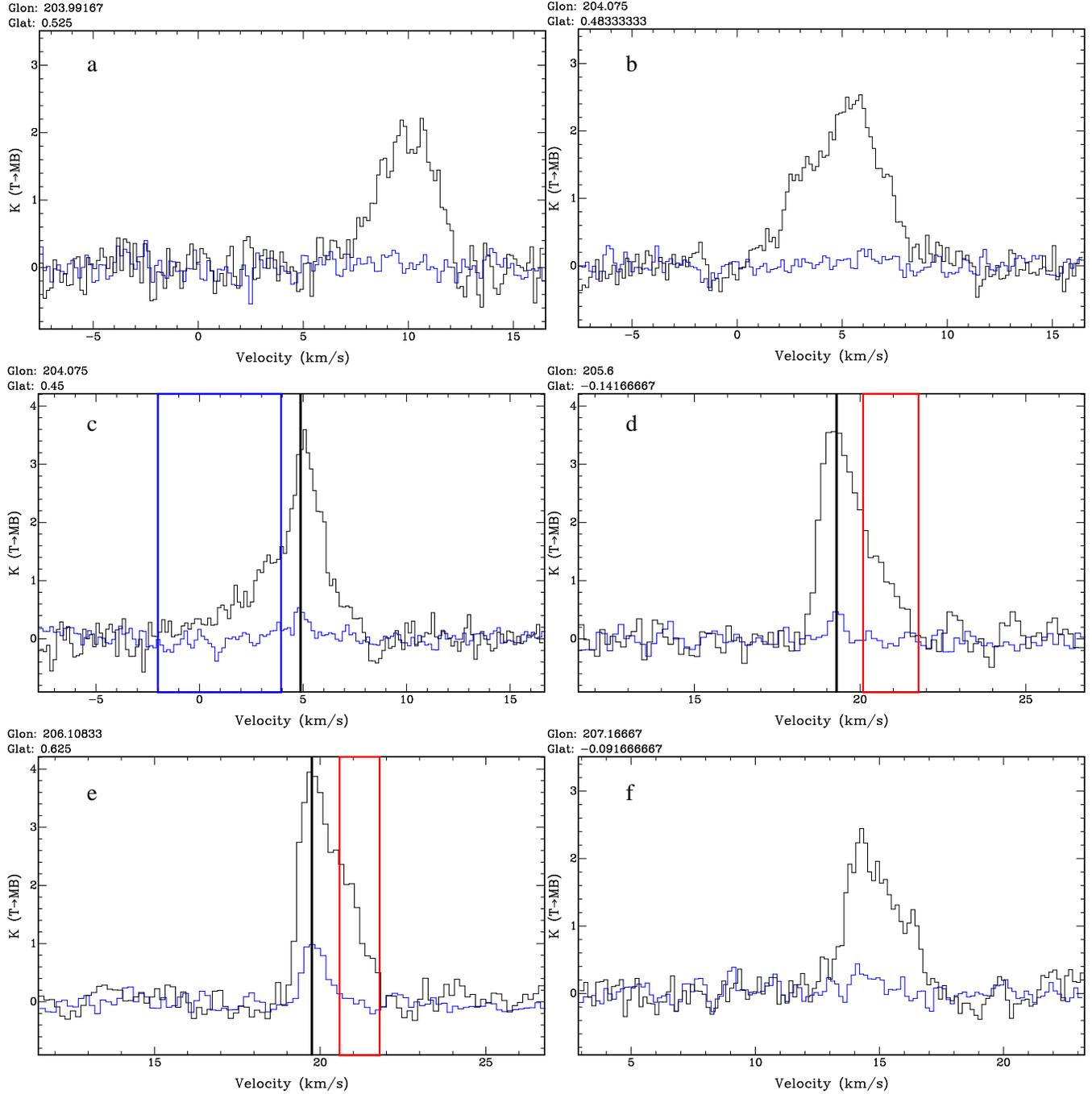}
\caption{
\twCO\ ($J$=1--0; black) and \thCO\ ($J$=1--0; blue) spectra of
shocked gas toward SNR G205.5$+$0.5. Positions of shocked MCs 
are labeled in Figure \ref{Mon12}.
These spectra are extracted from regions of 1$\times$1, 
1.5$\times$1.5, 1.5$\times$1.5, 1.5$\times$1.5, 2$\times$2, and 
2$\times$2 arcmin$^2$ for points a--f, respectively. The solid lines indicate 
the LSR velocity of unperturbed clouds from the \thCO\ peak. The red and 
blue rectangles show the velocity range of the wing from \twCO\ emission, 
which indicates the shocked gas.
Note that only selected typical spectra with relatively prominent 
asymmetric line structures are shown in the Figure.
\label{Monspec}}
\end{figure*}

Third, in the western edge of the SNR, an interesting MC structure 
is found to be physically associated with the remnant (see the 
right rectangle in Figure \ref{Mon12}). Figure \ref{Monsub}
displays striking spatial coincidence between the 
$V_{\rm LSR}\sim$5$\km\ps$ MC structure (in the interval of 3--12$\km\ps$) 
and the remnant's shell in the west boundary of the SNR.
In the left panel of this figure, a bright optical filament, 
as well as the radio ridge seen in 21 cm data, is elongated 
along the northeast--southwest direction. Note that both of the 11 cm 
\citep[the Effelsberg 11 cm survey, see][]{1990A&AS...85..691F} and the 6 cm 
\citep[the Parkes-MIT-NRAO (PMN) 4850 MHz survey, see][]{1993AJ....106.1095C} 
radio emission also can be 
discerned from the ambient background emission in such a region
with higher resolution and sensitivity. 
The southeastern boundary of the 
MC is very sharp and follows exactly the remnant's shell and
resembles the outline of the bright optical filaments and the radio 
ridge. That is, the molecular gas forms a wall facing the bright 
optical filaments or the SNR's shell. We found line-broadening 
features in such molecular gas (e.g., see a--c in 
Figure \ref{Monspec}). Especially, a narrow component centered 
at 5$\km\ps$ in the \thCO\ line, together with a blue wing component 
down to $-2\km\ps$ in \twCO\ line, is clearly recognized at 
position c, which can also be confirmed from Figure \ref{Monpv}. 
It seems that the shocked gas traced by the line 
broadening is near the region of the \twCO\ intensity maximum 
(see the red contours in the left panel of Figure \ref{Monsub}).
Furthermore, a substantial velocity gradient of the molecular 
gas (Figure \ref{Monpv}), is found to be mostly orthogonal to 
the shell of the SNR (refer to the direction denoted by the 
arrow in Figure \ref{Monsub}). The velocity gradient of 
$\sim0.9\km\ps$arcmin$^{-1}$ 
(or $\sim1.9\km\ps$pc$^{-1}$ at 1.6 kpc)
is very likely related to the SNR's shock.

\begin{figure*}
\includegraphics[trim=0mm 0mm 0mm 80mm,scale=0.87,angle=0]{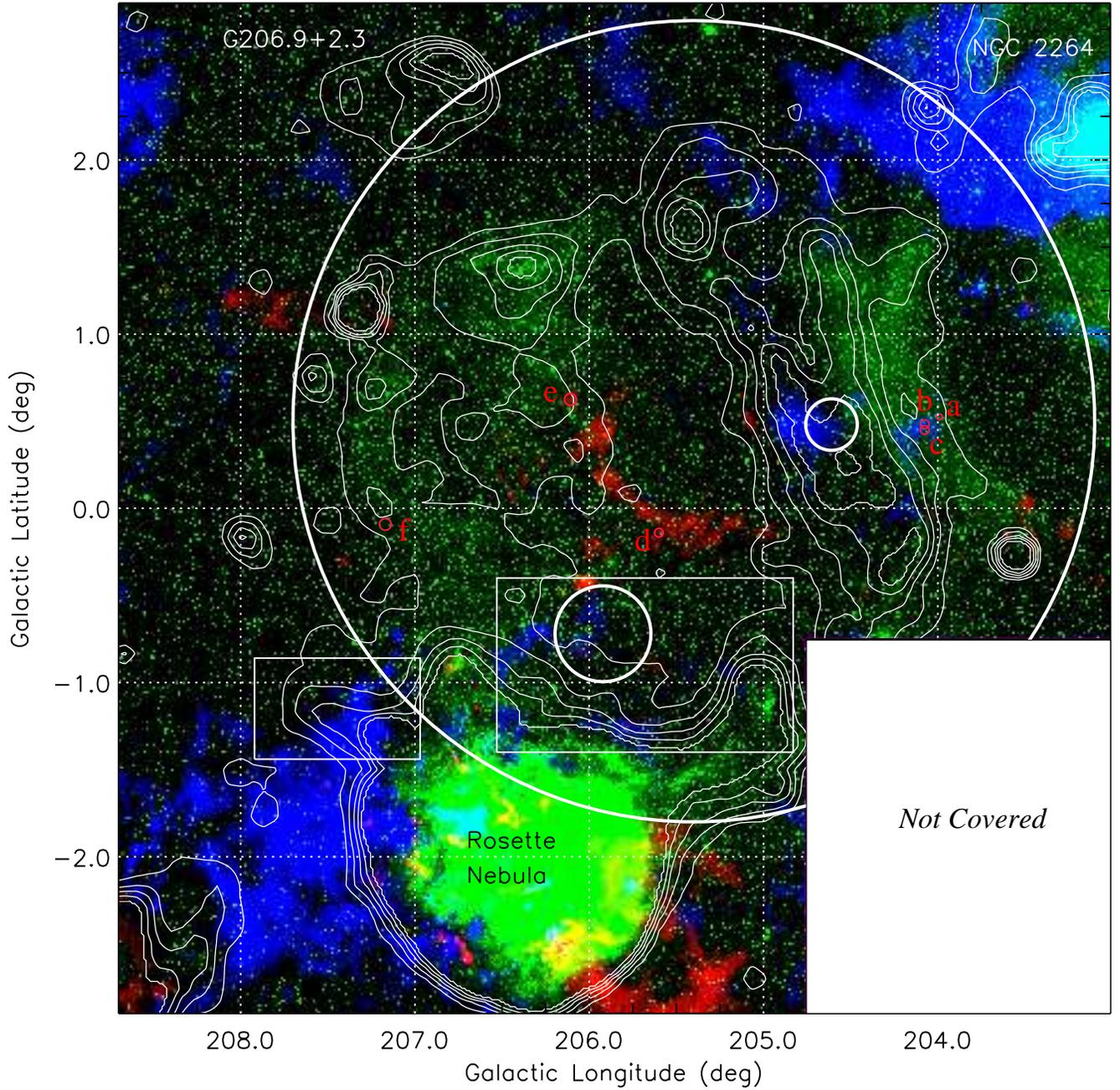}
\caption{
Integrated \twCO\ ($J$=1--0) emission toward SNR G205.5$+$0.5
in the interval of 3--12~km~s$^{-1}$ (blue) and 18--23~km~s$^{-1}$ (red), 
overlaid with the same radio contours as in Figure \ref{Mon12}.
The optical emission is shown in green.
The two rectangles indicate regions shown in Figures \ref{Monsm36} 
and \ref{Monsm27295}, respectively. 
Sizes of the red circles are enlarged to two times with respect to 
their angular sizes (1$'$--2$'$).
The large white circle indicates the best-fit Gaussian spatial model
from $FERMI/LAT$ observations after subtracting the background and the
emission from the Rosette Nebula \citep{2016ApJ...831..106K},
while the two little white circles show the 2FGL sources of J0636.0$+$0554
and J0637.8$+$0737 \citep{2012ApJS..199...31N}.
\label{Mongas}}
\end{figure*}

We thus suggest that the MCs at $\sim5\km\ps$ and $\sim19\km\ps$ 
are both associated with the remnant.
Figure \ref{Mongas} shows the gas distribution of the two components.
The 5$\km\ps$ molecular gas (blue) lies in front of the remnant and 
has been accelerated toward us, which can be confirmed from the high 
optical obscuration (the left panel of Figure \ref{Monsub}) and the 
molecular gas's blueshifted profile in the direction (e.g., position 
c in Figure \ref{Monspec}). The 19$\km\ps$ gas (red), however, is 
probably on the farside of the remnant because of their little 
evidence of associated LOS obscuration and the 
redshifted profiles (see asymmetric line-broadening features at
positions d--f in Figure \ref{Monspec}). It is worth
mentioning that not all molecular gas in Figure \ref{Mongas} 
is interacting with SNR G205.5$+$0.5. For example, the bule part
in the upper-right corner of Figure \ref{Mongas}, which belongs to
the NGC 2264 GMC complex and is located at a distance of $\sim740$--900~pc 
\citep[e.g.,][]{1997AJ....114.2644S,2009AJ....138..963B,2014ApJS..211...18K} 
is probably not related to the SNR.

The excellent spatial correlation between the ionized gas traced 
by optical data and the molecular gas traced by CO data, as well as 
broad wings of line profiles in some regions, suggests that SNR 
G205.5$+$0.5 is interacting with the ambient molecular gas.
We note that the nearby Rosette Nebula is also interacting
with the same CO component, which is consistent with the result
of other studies
\citep[e.g.,][]{1998A&A...335.1049S,2009MNRAS.395.1805D}.
In Figure \ref{Mongas}, the Rosette Nebula is found to be surrounded
by the 3--12$\km\ps$ and 18--23$\km\ps$ MCs, indicating a connection
between them. A detailed study 
between the Rosette Nebula and its ambient MCs will be presented in a 
forthcoming paper (C. Li et al. 2017, in preparation).

Furthermore, based on the 3D extinction map from 
\cite{2015ApJ...810...25G}, we can estimate the distance toward the
molecular gas associated with SNR G205.5$+$0.5 to constrain the distance of the
SNR. We select a region centered at ($l=$204\fdg107, $b=$0\fdg471),
which is just located at the CO emission peak of the 5$\km\ps$ MCs
(see the maximum of red contours in the left panel of Figure \ref {Monsub}). We find a
rapid increase for this region at a distance modulus (DM) of $\sim$11
($\sim$1.6 kpc). On the other hand, several regions are selected for the
19$\km\ps$ MCs, which are at DM$\sim$11--11.5 ($\sim$1.6--2 kpc).
Since the 5$\km\ps$ and 19$\km\ps$ MCs are both physically associated with the
SNR, it indicates that SNR G205.5+0.5 is very likely located at a distance of
$\sim$1.6 kpc. This value is coincident within the
errors with a distance of 1.6 kpc according to the association 
between the SNR and the Rosette Nebula \citep[refer to works of distance
estimate of the Nebula; e.g.,][]{2000A&A...358..553H,2002AJ....123..892P}.

In combination with the above analysis, we suggest that SNR
G205.5$+$0.5 and the nearby Rosette Nebula are associated and both
of the two extended sources are at a similar distance of 1.6 kpc.
Our new CO analysis confirms the speculation that SNR G205.5$+$0.5 
is probably associated with the nearby Rosette Nebula
\citep[e.g.,][]{1996A&A...315..578O}.
Using the CfA 1.2 m millimeter-wavelength telescope, \cite{1996A&A...315..578O}
found that MCs 18-24 (see Tables 1 and 2, Figures 2 and 6 in their paper)
lie toward the two extended sources and are probably related to their nearby
bright optical and radio emission.
They estimated the emission-weighted
average kinematic distance of 1.6 kpc for these MCs, which is in agreement
with the photometric distance to the Rosette Nebula. Recently, 
\cite{2012A&A...545A..86X} revealed partial neutral hydrogen shell structures
outside the western boundary of the remnant. The interesting HI shell in
the velocity interval of 5--25$\km\ps$ seems to relate to SNR G205.5$+$0.5 
and the Rosette Nebula, which strengthens the association between them.

Using the relation of initial total energy $E$ and the 
ambient gas density $n_{0}$ 
\citep[e.g., $E_{51}=5.3\times10^{-8}n_{0}^{1.12}({\rm cm}^{-3})v_{\rm sh}^{1.40}({\rm km~s^{-1}})R^{3.12}({\rm pc})$~erg,
Equation 26 in][]{1974ApJ...188..501C},
the ambient density of the remnant is $\sim0.4E_{51}^{0.893}$~cm$^{-3}$ 
for a remnant's radius of $\sim$51~pc (e.g., size of $\sim220'$ at a 
distance of 1.6~kpc) and an expansion velocity of $\sim50\km\ps$ 
\citep{1976ApJ...207...53W,2001A&A...372..516W,2016ApJ...819...45D}, 
where $E_{51}$ is the initial total energy of the remnant in units of 
10$^{51}$~ergs. The low shock velocity of $\sim50\km\ps$ is consistent 
with the result that the remnant has weak [O $\textsc{iii}$] emission 
\citep{1985ApJ...292...29F}. If the remnant is expanding into the ambient 
gas with such velocity, its age is estimated to be $\sim3.0\times10^5$~years 
based on the radiative-phase solution 
\citep[e.g., $R\sim t^{0.305}$, Equation 25 in][]{1974ApJ...188..501C}.

An extended or confused source, 3EG J0634$+$0521 
\citep{1999ApJS..123...79H,2008A&A...489..849C}, and a TeV source,
HESS J0632$+$057 \citep{2007A&A...469L...1A}, seem to be located
close to the interface between SNR G205.5$+$0.5 and the Rosette
Nebula (see blue circles in Figure \ref{Mon12}). However, there is
no further evidence to show the association between the two 
high-energy sources and the SNR. On the other hand, 
2FGL sources of J0636.0$+$0554 and J0637.8$+$0737 
\citep{2012ApJS..199...31N} are located close to the molecular 
gas revealed by our new CO survey (two little circles in Figure \ref{Mongas}).
Very recently, extended gamma-ray emission from $FERMI$/LAT observations was 
found inside the remnant 
\citep[see Figures 3 and 4 in][]{2016ApJ...831..106K}.
According to our study, the MC concentration in the velocity
intervals of 3--12$\km\ps$ and 18--23$\km\ps$, which is shown to be
associated with the remnant, is consistent well with the 
distribution of the background-subtracted gamma-ray emission.
The confirmation of SNR--MC interaction toward SNR G205.5$+$0.5, 
as well as the spatial coincidences between the shocked molecular 
gas and the gamma-ray emission, indicates that the extended gamma-ray 
emission inside the remnant is very likely from the hadronic scenario.
In the scenario, the high-energy emission near the old SNR can
be naturally explained by the decay of neutral pions produced in interactions 
between hadrons accelerated by the remnant's shock and its ambient dense 
molecular gas, which is also consistent with the theoretical analysis 
\citep[e.g.,][]{2006MNRAS.371.1975Y,2009ApJ...707L.179F}.

\subsection{SNR G206.9$+$2.3 (PKS 0646$+$06)}
\begin{figure*}
\includegraphics[trim=5mm 0mm 0mm 150mm,scale=0.9,angle=0]{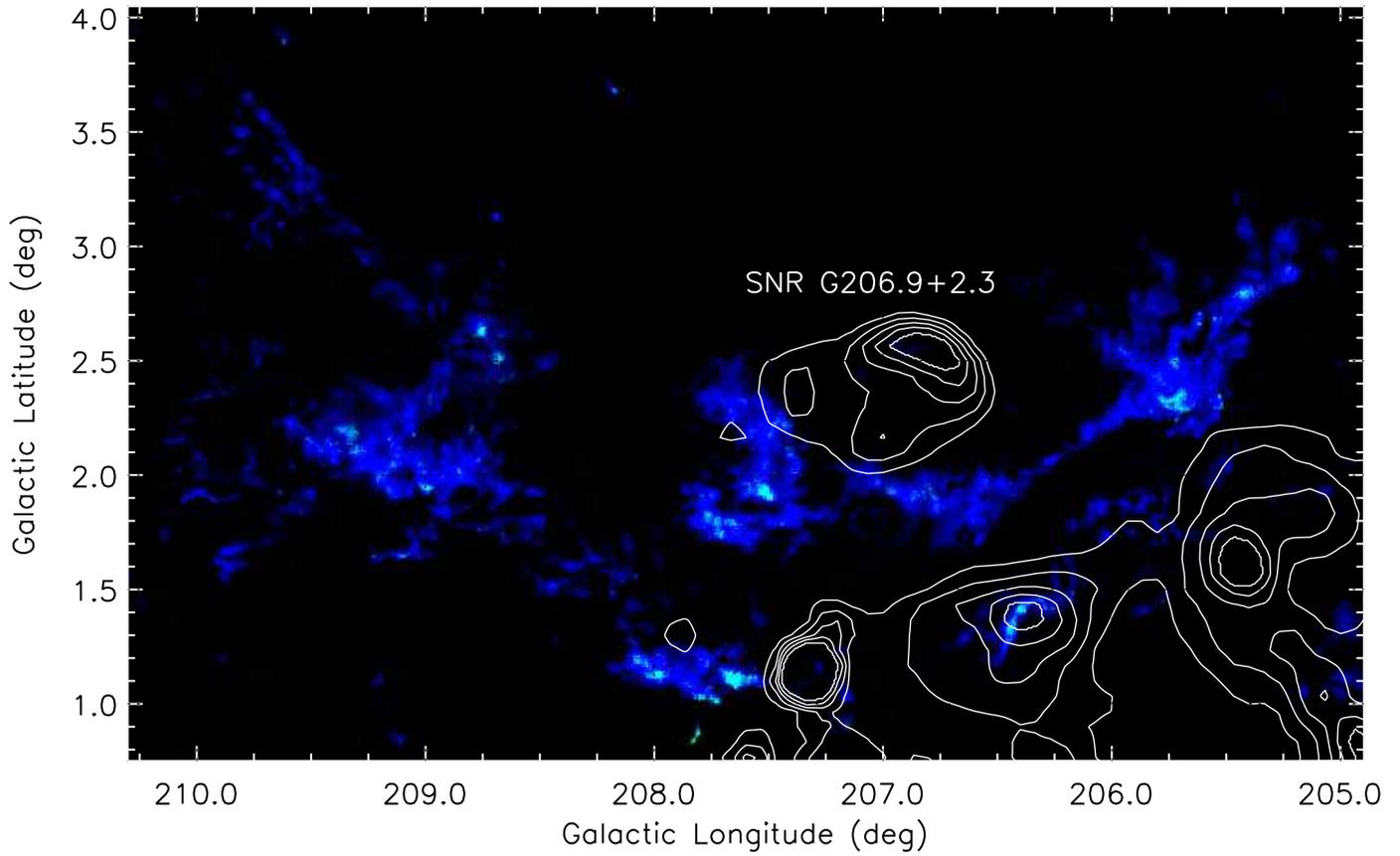}
\caption{
\twCO\ ($J$=1--0, blue) and \thCO\ ($J$=1--0, green) intensity map 
toward G206.9$+$2.3 in the 11--19~km~s$^{-1}$ interval,
overlaid with radio continuum contours from the Effelsberg 21 cm survey
\citep{1997A&AS..126..413R}.
The radio contours levels are at 340, 480, 620, 760, and 900 mK.
\label{G207color}}
\end{figure*}

SNR G206.9$+$2.3, which is also known as PKS 0646$+$06,
is a faint and extended nonthermal radio source near the 
Monoceros Nebula (Figures \ref{Mon12} and \ref{G207color}).
The radio spectral index of the SNR is about $-0.45$
\citep{1982A&A...109..145G}, which was recently confirmed
by \citet[i.e., $\alpha=-0.47$]{2011A&A...529A.159G}.
The remnant, which has relatively strong radio emission
in its northwestern region, roughly shows an elliptical shape
in radio emission (see contours in Figure \ref{G207color}). 
Near the northwestern radio peak, the optical data reveals
a bright thin filament with a relatively high density 
\citep[e.g., $n_{\rm e}\sim800$~cm$^{-3}$, position 1 
at $l=$206\fdg77 and $b=$2\fdg61 in][]{1985ApJ...292...29F}. 
The enhanced X-ray emission from Einstein observations, which 
is close to the northwestern radio peak of the remnant, is 
suggested to be associated with SNR G206.9$+$2.3 \citep{1986A&A...156..191L}.
Both of the X-ray and optical studies indicate that the SNR
is probably evolving in low-density environment 
\citep{1986A&A...156..191L,2014RMxAA..50..323A} except for the 
relatively high density region near the radio peak (e.g., 
position 1 mentioned above).

\begin{figure*}
\includegraphics[trim=5mm 0mm 0mm 150mm,scale=0.95,angle=0]{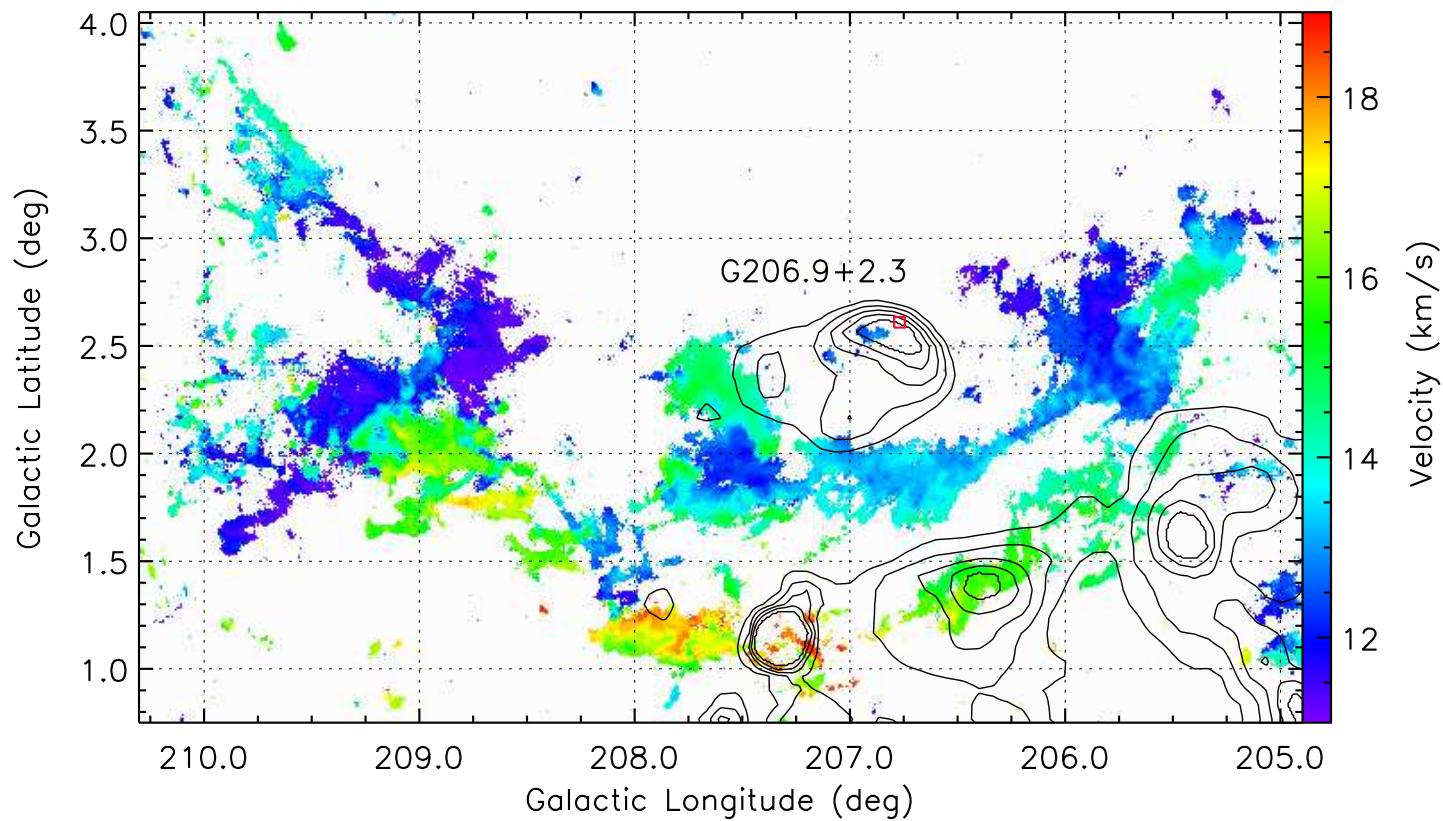}
\caption{
Intensity-weighted \twCO\ ($J$=1--0) mean velocity (first moment) map 
of MCs toward SNR G206.9$+$2.3 in the interval of 11--19~km~s$^{-1}$,
overlaid with the same radio contours as in Figure \ref{G207color}.
The red box indicates position 1 in \cite{1985ApJ...292...29F}.
\label{G207vel}}
\end{figure*}

We covered a large CO map toward SNR G206.9$+$2.3 and the nearby 
regions with high sensitivity and resolution. CO emission is very 
weak in the direction of the remnant except for the nearby MC 
complex at $V_{\rm LSR}\sim$15$\km\ps$ (in the velocity interval 
of 11--19$\km\ps$, see Figure \ref{G207color}). The main body of 
the MC complex, which has weak \thCO\ emission over the FOV 
(see green parts in the figure), displays curved partial shell 
structures and extends over 5 degrees in longitude from 
204\fdg9 to 210\fdg3. We find that the MC complex has a complicated 
velocity structure (Figure \ref{G207vel}). The 13--15 $\km\ps$ 
molecular gas, together with the outside large-scale partial shell 
structures at 15--17$\km\ps$, seems to form a molecular void toward 
the SNR. The curvature of the partial shell structures 
points in the direction from the SNR to the outside 
shell-like MCs. It is interesting to note that SNR G206.9$+$2.3 
roughly lies in the geometrical center of the MC's void, which 
agrees well with the low extinction values in the direction 
\citep[see the case of G206.9$+$2.3 in Table 2 of][]{1985ApJ...292...29F}. 
Some patches of molecular gas at an LSR velocity of $\sim$13$\km\ps$ 
are found to coincide with the northwestern radio peak of 
the remnant (e.g., $l\sim$206\fdg85 and $b\sim$2\fdg55, near 
the red box shown in Figure \ref{G207vel}), which is consistent 
with the fact of the relatively high density there based on 
optical observation \citep[refer to position 1 of SNR 
G206.9$+$2.3 in][]{1985ApJ...292...29F}. 

\begin{figure*}
\includegraphics[trim=5mm 0mm 0mm 150mm,scale=0.95,angle=0]{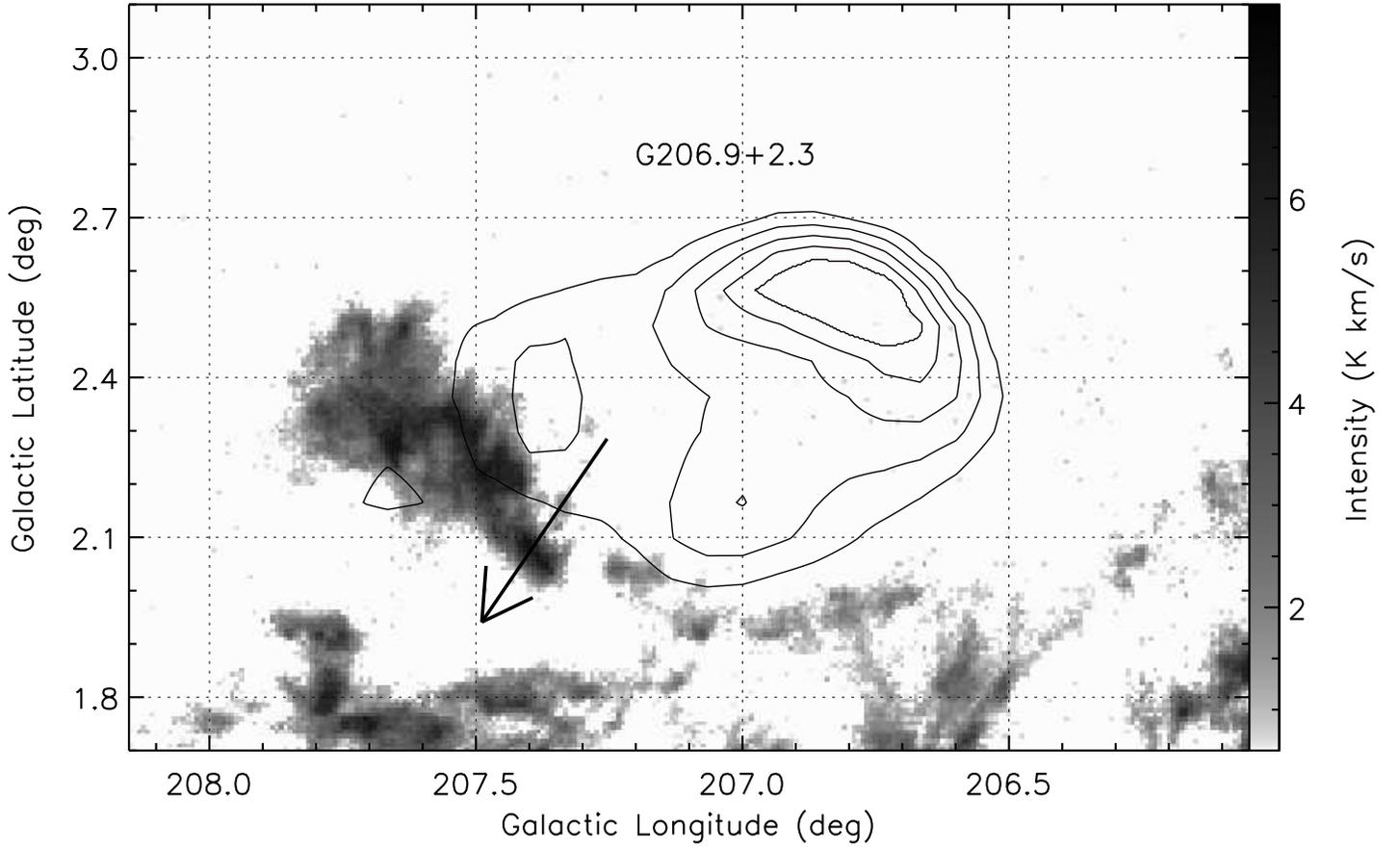}
\caption{
Integrated \twCO\ ($J$=1--0) emission toward SNR G206.9$+$2.3 from
14 to 15~km~s$^{-1}$, overlaid with the same contours as in 
Figure \ref{G207color}. The square-root scale is used to enhance 
the faint CO emission along the southern boundary of the remnant.
\label{G207shell}}
\end{figure*}

\begin{figure*}
\includegraphics[trim=-110mm 0mm 0mm -10mm,scale=0.4,angle=0]{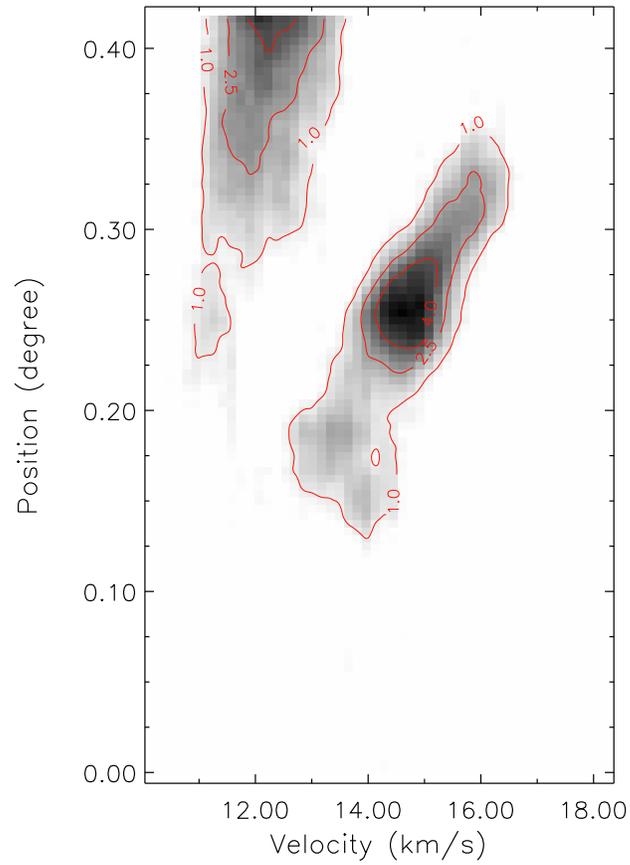}
\caption{
PV diagram of \twCO\ ($J$=1--0) emission in the southeast
of SNR G206.9$+$2.3. The PV slice with length of 25$'$ 
(from ($l=$207\fdg254, $b=$2\fdg285) to ($l=$207\fdg489, $b=$1\fdg941))
and width of 4\farcm5 is perpendicular to the SNR's radio shell and the MC partial 
shell structures (see the arrow in Figure \ref{G207shell}).
The red contours indicate the emission of \twCO\ in units of K.
\label{G207pv}}
\end{figure*}

In particular, the 13--15 $\km\ps$ molecular gas is well along 
the southeastern boundary of the SNR. The close-up partial shell 
structure in the velocity interval of 14--15$\km\ps$ is presented 
in Figure \ref{G207shell}, in which the SNR is clearly surrounded 
by molecular shell from the east to the south. Moreover, a 
velocity gradient of $\sim0.3\km\ps$arcmin$^{-1}$ 
(or $\sim0.6\km\ps$pc$^{-1}$ at 1.6 kpc, see the PV diagram 
in Figure \ref{G207pv}) is detected to be perpendicular to (see 
the arrow in Figure \ref{G207shell}) the shell structure, which 
probably indicates that the molecular gas is expanding outward 
at several kilometers per second. 

If SNR G206.9$+$2.3 is associated with the $\sim$15$\km\ps$ molecular
gas cavity, the kinematic distance of the remnant is estimated to be 
$\sim$1.6~kpc based on the A5 rotation curve model of \cite{2014ApJ...783..130R}. 
The remnant's angular size of $\sim60'$ thus corresponds to a 
physical radius of $\sim14.0$ pc.
Using the relation of the blast-wave energy $E$ and the cloud 
parameters $n_c$ and $v_c$ 
\citep[e.g., $E_{51}=2\times10^{-9}\beta^{-1}n_{0}({\rm cm}^{-3})v_{\rm sh}^{2}({\rm km~s^{-1}})R^{3}({\rm pc})$ erg,
Equation 26 in][]{1975ApJ...195..715M}, we can roughly estimate the 
shock velocity of the remnant. Adopting the numerical factor $\beta$=1, 
explosion energy $E$=10$^{51}$~ergs, and the density n$_0\sim$0.1~cm$^{-3}$
\citep[the upper limit, see discussions in][]{1986A&A...156..191L}, 
the shock velocity of the remnant is greater than $\sim1000\km\ps$. 
Therefore, SNR G206.9$+$2.3 is probably in the Sedov-Taylor phase 
and the age of the remnant is estimated to be less than 4000 years.

Finally, it should be noted that, although the above morphological 
correspondences may indicate an association between the SNR and the 
15~$\km\ps$ molecular gas, there are few kinematic signatures to 
suggest an interaction scenario directly (e.g., no significant line 
broadening is detected within or near the remnant). It probably 
indicates that the SNR is evolving in the molecular cavity, which is 
in agreement with the low density there and the surrounding molecular 
shell structures. Further observations and analysis of the remnant 
are required to draw any firm conclusions.

\subsection{SNR G213.0$-$0.6}
Using the 863 MHz data and the 2.695 GHz data from the Effelsberg
survey, \cite{2003A&A...408..961R} firstly identified G213.0$-$0.6 as
an old SNR with $\alpha\sim-0.4$. The remnant displays partial 
shell-like structures on a large scale and it has low radio surface 
brightness. On the other hand, a large complex of \HII\ region 
Sh 2-284, which has bright optical, infrared, and radio emission,
is located to the southwestern boundary of the remnant.
The remnant was also named as G213.3$-$0.4 according to the new 
optical H$\alpha$ imaging and relatively high-resolution radio 
observations from the PMN survey at 4.85 GHz and 
NRAO/VLA Sky Survey (NVSS) at 1.4 GHz \citep{2012MNRAS.419.1413S}. 
Following Green's SNR catalog \citep{2014BASI...42...47G}, we use 
SNR G213.0$-$0.6 as its name throughout this paper.
Generally speaking, the SNR is less studied because of its large angular 
size and low radio surface brightness. 

\begin{figure*}
\includegraphics[trim=5mm 0mm 0mm 90mm,scale=0.9,angle=0]{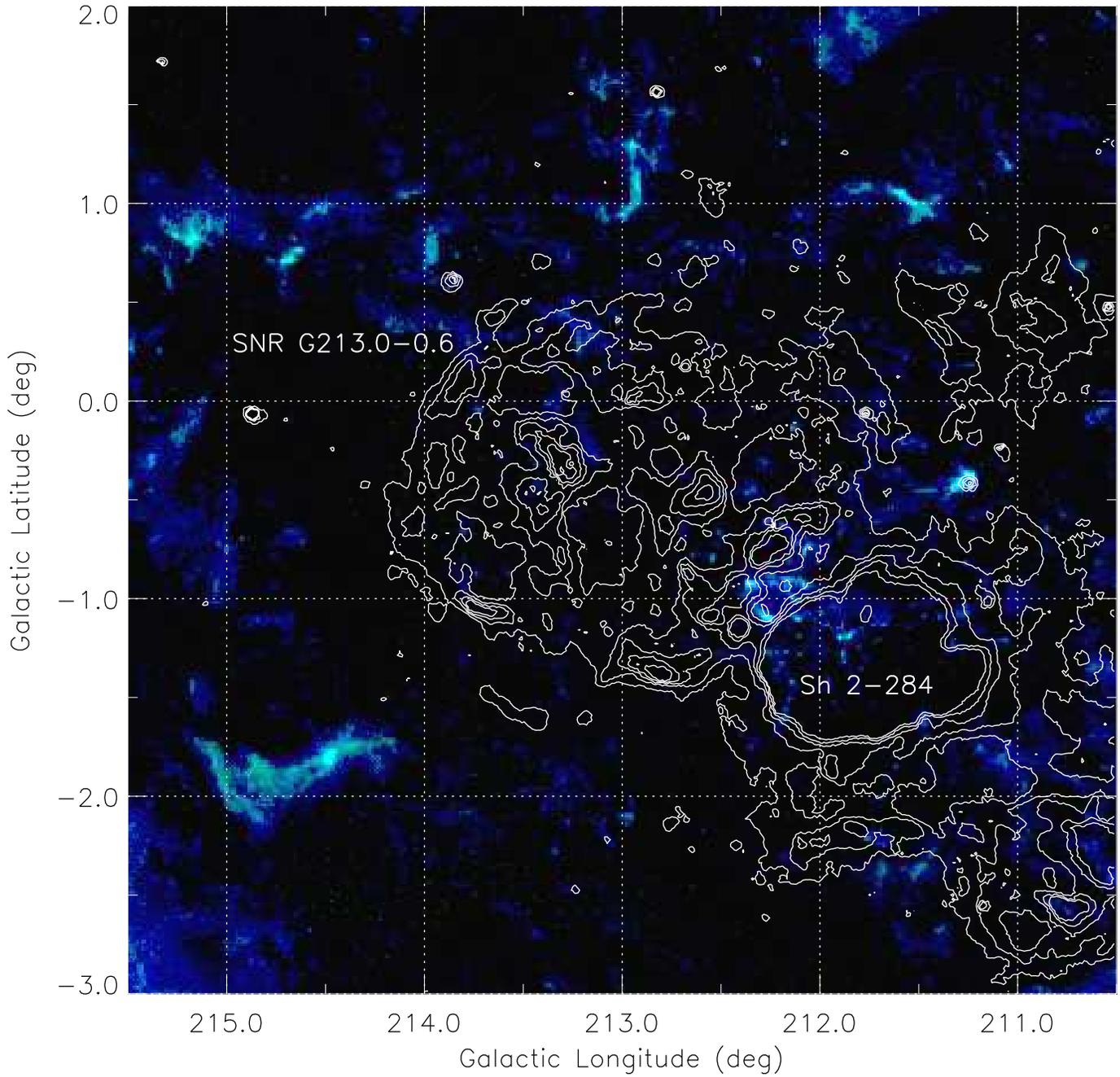}
\caption{
\twCO\ ($J$=1--0; blue) and \thCO\ ($J$=1--0; green) intensity map 
toward SNR G213.0$-$0.6 in the 0--60~km~s$^{-1}$ interval,
overlaid with optical contours from the Southern H$\alpha$ 
Sky Survey \citep{2005MNRAS.362..689P}. 
The SHASSA contour levels are at 220, 296, 372, 448, and 524 decirayleighs.
The \HII\ region Sh 2-284 is also labeled in the lower-right corner.
\label{G213rgb}}
\end{figure*}

Figure \ref{G213rgb} shows the distribution of the molecular gas
toward SNR G213.0$-$0.6. 
Given the very fragmented structures of the SNR in 
radio emission, optical contours from the 
Southern H$\alpha$ Sky Survey \citep{2005MNRAS.362..689P}
have been made to trace the SNR and the southwestern \HII\ 
region Sh 2-284.
In the direction of SNR G213.0$-$0.6, 
molecular gas is mainly concentrated in the velocity
range of 0--60$\km\ps$. CO emission is very weak toward 
SNR G213.0$-$0.6, while the emission outside the remnant is 
relatively strong. This fact probably corresponds to the faint radio
emission on a large scale because the density of the SNR's environment 
is relatively low.
In Figure \ref{G213rgb}, the southwestern 
molecular gas is obviously related to \HII\ region Sh 2-284. 
The relationship between the SNR and its
southwestern extended source Sh 2-284 is also investigated.

\begin{figure*}
\includegraphics[trim=-5mm 0mm 0mm 100mm,scale=0.9,angle=0]{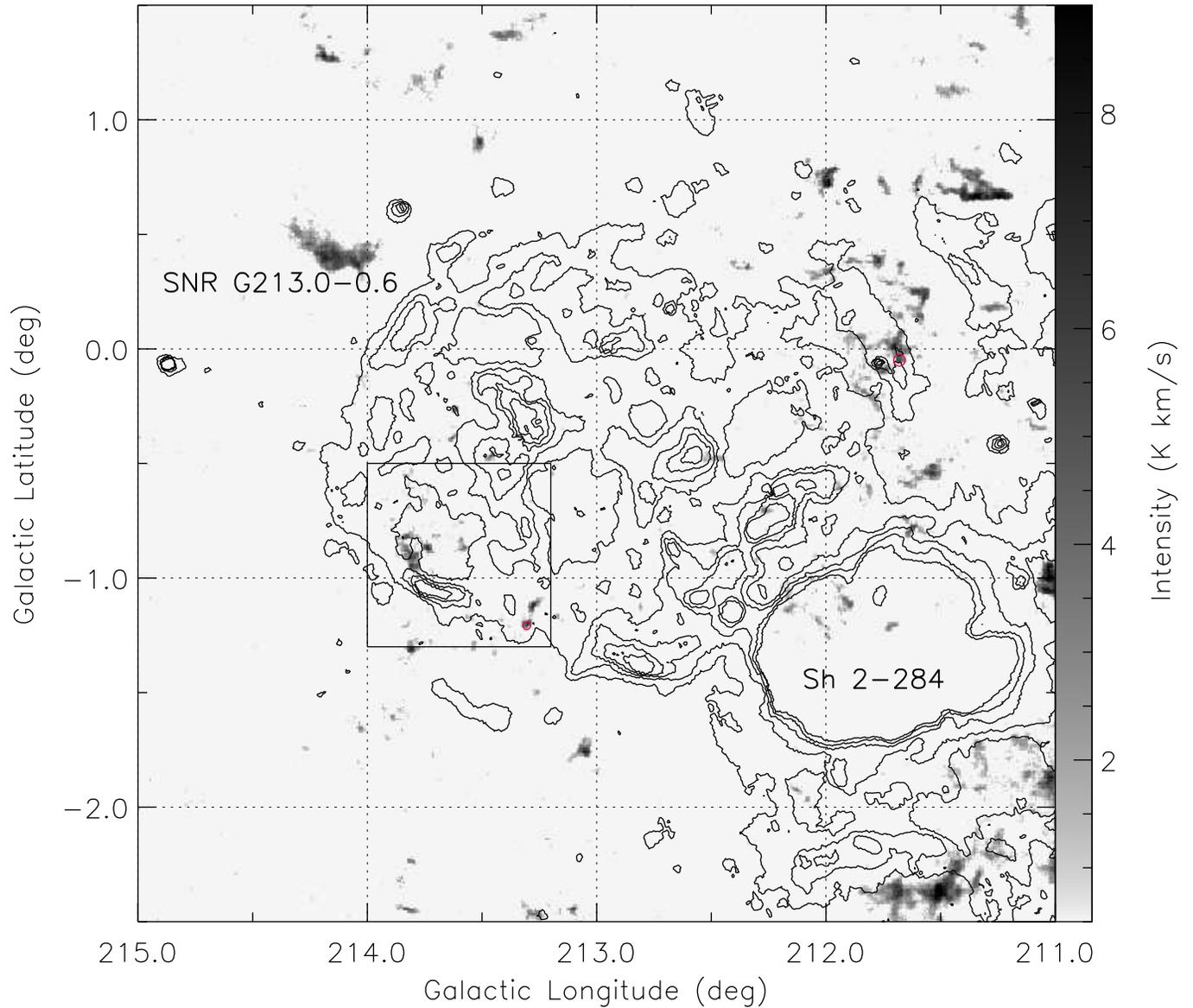}
\caption{
Integrated \twCO\ ($J$=1--0) emission toward SNR G213.0$-$0.6 from 2 to 
15~km~s$^{-1}$, overlaid with the same contours as in Figure \ref{G213rgb}. 
The box indicates the region shown in Figure \ref{G213velg}.
Two red circles indicate positions of shocked gas, of which spectra
are shown in Figure \ref{G213spec}. 
\label{G213main2}}
\end{figure*}

\begin{figure*}
\includegraphics[trim=-20mm 0mm 0mm 110mm,scale=0.75,angle=0]{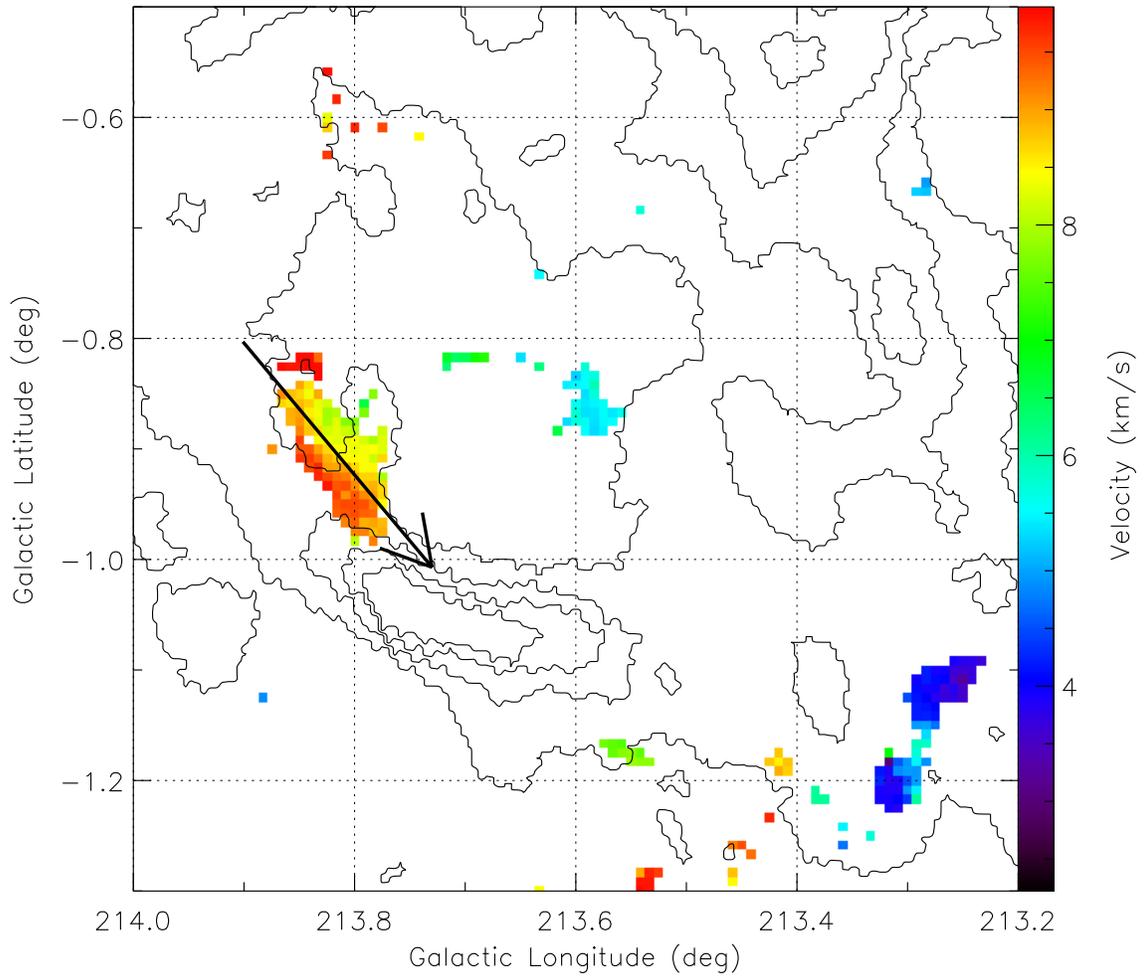}
\caption{
Intensity-weighted \twCO\ ($J$=1--0) mean velocity (first moment) map 
of MCs in the interval of 2--10~km~s$^{-1}$,
overlaid with the same contours as in Figure \ref{G213rgb}.
The arrow indicates the PV slice shown in Figure \ref{G213longpv}.
\label{G213velg}}
\end{figure*}

\begin{figure*}
\includegraphics[trim=-90mm 0mm 0mm 0mm,scale=0.5,angle=0]{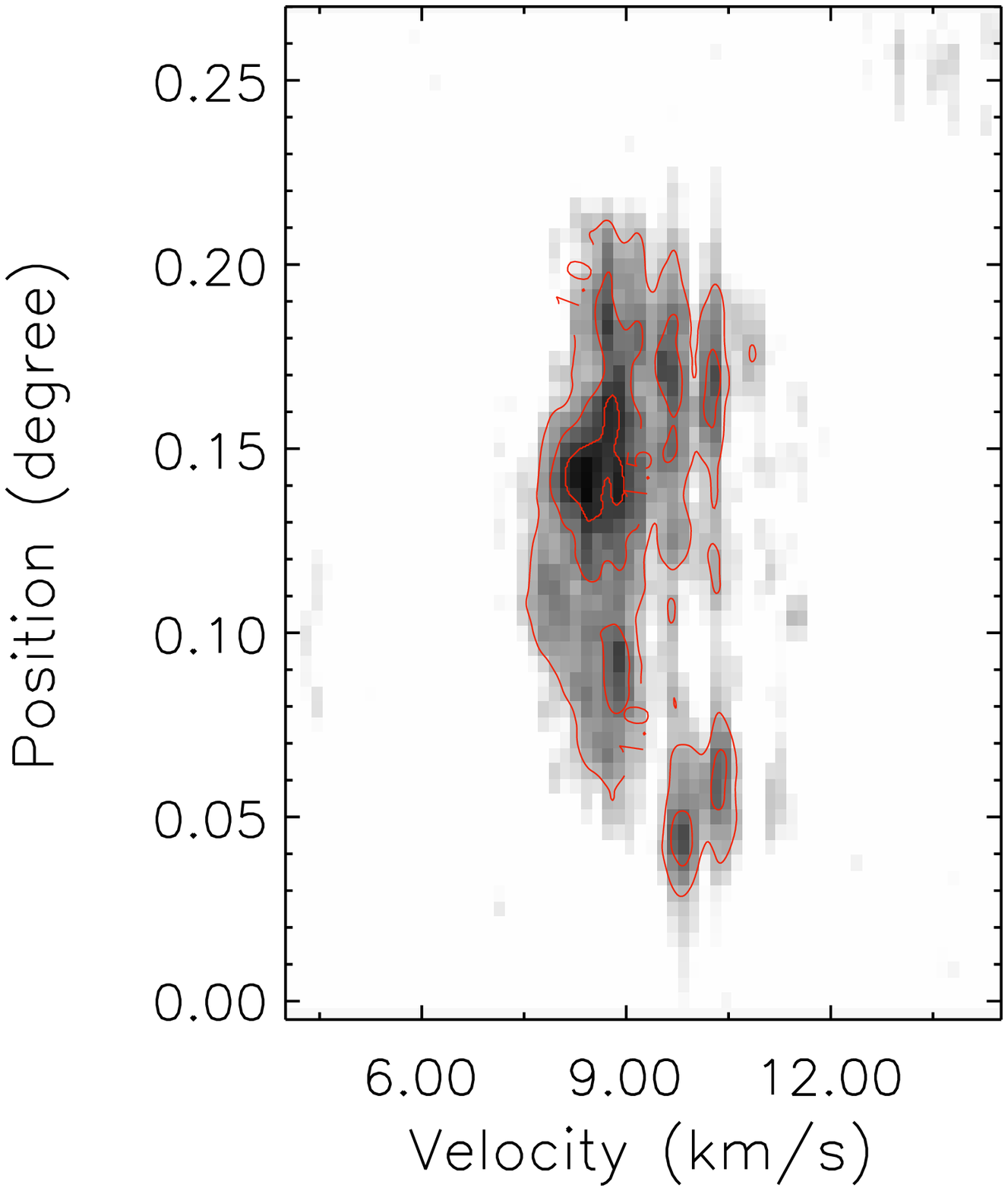}
\caption{
PV diagram of \twCO\ ($J$=1--0) emission near the H$\alpha$ peak of 
SNR G213.0$-$0.6's southeastern boundary
\citep[see Figure 5a in][]{2012MNRAS.419.1413S}.
The PV slice with a length of 16$'$ 
(from ($l=$213\fdg901, $b=-$0\fdg803) to ($l=$213\fdg730, $b=-$1\fdg007))
and a width of 4\farcm5 is along the SNR's
radio shell (see the arrow in Figure \ref{G213longpv}).
The red contours indicate the emission of \twCO\ in units of K.
\label{G213longpv}}
\end{figure*}

We find that molecular gas at systemic velocity of
8--11$\km\ps$ seems to be located around the boundary of SNR G213.0$-$0.6, especially
to the southeast and northwest (Figure \ref{G213main2}).
In projection, the H$\alpha$ and radio emission is enhanced
toward the southeastern boundary of the remnant (see the box
region in Figure \ref{G213main2}) coincident with an MC with
a velocity interval of 6.5--12$\km\ps$. The MC, which is found to 
be along the boundary of the remnant from the northeast to the southwest,
displays a prominent velocity gradient ($\sim0.7\km\ps$arcmin$^{-1}$,
or $\sim2.4\km\ps$pc$^{-1}$ at 1.0 kpc, see below) perpendicular to
the main axis of the \twCO\ emission (see the arrow in Figure \ref{G213velg}).
The PV diagram of the molecular gas (Figure \ref{G213longpv}) 
reveals the detailed velocity structure along the MC, as well as 
an interesting arc structure seen in the map.  
We suggest that this feature is probably related to the remnant's shock. 
\cite{2012MNRAS.419.1413S} found that the density in 
the region of the H$\alpha$ peak
is relatively high based on optical observations 
($n_{\rm e}\sim10^{2}$~cm$^{-3}$, refer to Table 2 in their paper). 
This result suggests that SNR G213.0$-$0.6 is interacting with 
a clumpy interstellar medium, which agrees with the 
CO emission near the H$\alpha$ peak using our CO data 
(Figure \ref{G213velg}). Assuming a constant value of $X_{\rm CO}$=
$2\E{20}$~cm$^{-2}$K$^{-1}$km$^{-1}$s \citep{2013ARA&A..51..207B}, 
the MC's thickness of 4$'$, and a distance of 1~kpc (see below), 
the density of the MC is about 300$d_{1\rm {kpc}}^{-1}$~cm$^{-3}$,
which is consistent with the result of the optical
study \citep{2012MNRAS.419.1413S}.
The interesting MC, as well as its velocity structure, 
suggests that the molecular gas with LSR velocity $\sim$8--11$\km\ps$ 
is probably interacting with SNR G213.0$-$0.6.

\begin{figure*}
\includegraphics[trim=5mm 0mm 0mm 0mm,scale=0.67,angle=0]{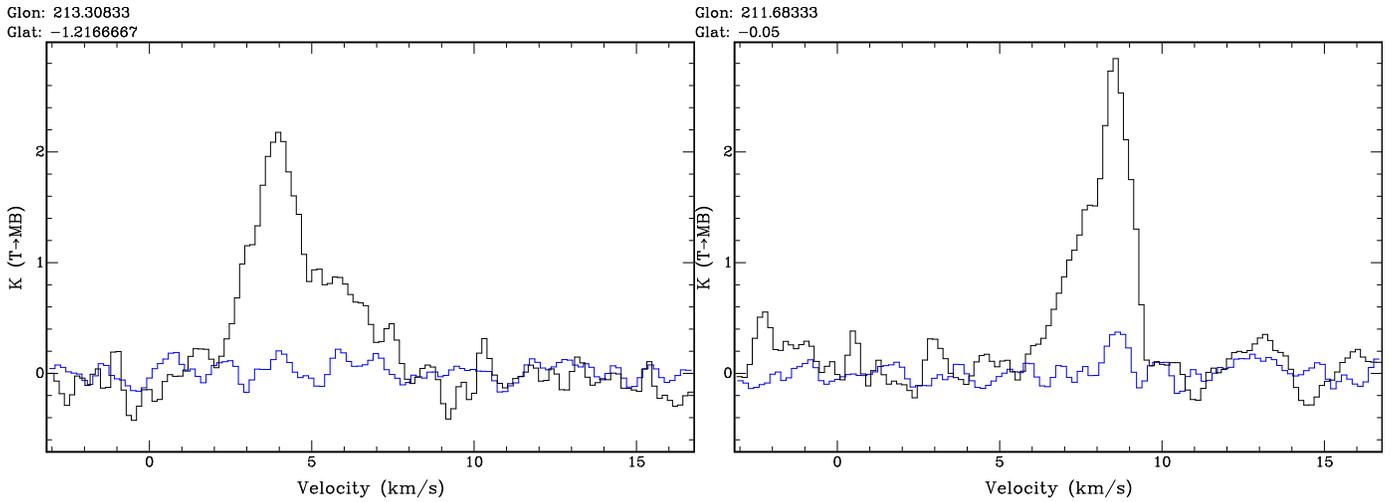}
\caption{
\twCO\ ($J$=1--0; black) and \thCO\ ($J$=1--0; blue) spectra of
the shocked gas toward SNR G213.0$-$0.6. The two positions of the 
shocked MCs are labels red circles in Figure \ref{G213main2}.
The two spectra are extracted from regions of 2$\times$2 and 3$\times$2 
arcmin$^2$ for the left panel and the right panel, respectively.
\label{G213spec}}
\end{figure*}

CO emission of this molecular component is also 
concentrated in the northwest of the remnant, in which the H$\alpha$ 
and radio emission displays blow-out structure (see the northwestern 
region in Figure \ref{G213main2}). An MC with line broadening (several 
$\km\ps$, see the right panel of Figure \ref{G213spec}) is revealed in 
the region (see the relatively large red circle in the northwest of 
Figure \ref{G213main2}). The H$\alpha$ emission, together with the
PMN 4850 MHz radio emission, appears to be locally enhanced 
in such a region.

In combination with the morphological 
correspondence and the broadening of CO profiles, we 
suggest that molecular gas at $V_{\rm LSR}\sim$9$\km\ps$
is physically associated with the remnant. Accordingly, the 
kinematic distance of the remnant is about 0.7--1.0 kpc (e.g., 
$V_{\rm LSR}\sim$8--11$\km\ps$ from Figure \ref{G213longpv})
based on the A5 rotation curve model of \cite{2014ApJ...783..130R}.
The estimated kinematic distance of the remnant is consistent 
well with that of the empirical radio surface-brightness-to-diameter 
($\Sigma$--$D$) relation. Using the updated $\Sigma$--$D$
relation for SNRs and the probability-density-function-based method
\citep{2014SerAJ.189...25P}, the distance of the SNR is estimated 
to be 1.0$\pm0.18$ kpc. 
We also estimate that the 8--11$\km\ps$ molecular gas 
(e.g., at position ($l=$213\fdg798, $b=-$0\fdg949), see Figure \ref{G213velg} ) 
is at DM$\sim$10 (or $\sim$1.0 kpc) according to the 3D extinction
map \citep{2015ApJ...810...25G}. 
The estimated distance is in good agreement with the result 
from the above $\Sigma$--$D$ and kinematic methods.
We thus adopt 1 kpc as the distance to SNR
G213.0$-$0.6. 

The angular size ($\sim$3\fdg4 $\times$ 2\fdg3, 
accounting for the blow-out structure in the northwestern region) 
of the remnant corresponds to a physical size of 59~pc $\times$ 40~pc 
at about 1 kpc distance. The large physical size, the low radio surface
brightness, and the relatively high density near the boundary of the
remnant indicates that SNR G213.0$-$0.6 probably exploded in a 
low-density cavity and now is interacting with the surrounding clumpy 
interstellar medium.
Assuming the mean density of $n_0\sim$0.1~cm$^{-3} \equiv n_{-1}$ and 
the SNR's explosion energy of $E\sim10^{51}$~erg $\equiv E_{51}$, 
the remnant's equivalent radius of 24.3~pc yields a shock velocity 
of about 590$(E_{51}/n_{-1})$$^{0.5}\km\ps$. The age of the remnant 
in the radiative phase is thus about 1.1$(n_{-1}/E_{51})$$^{0.5}\times10^4$ 
years.

\begin{figure*}
\includegraphics[trim=0mm 0mm 0mm 100mm,scale=0.9,angle=0]{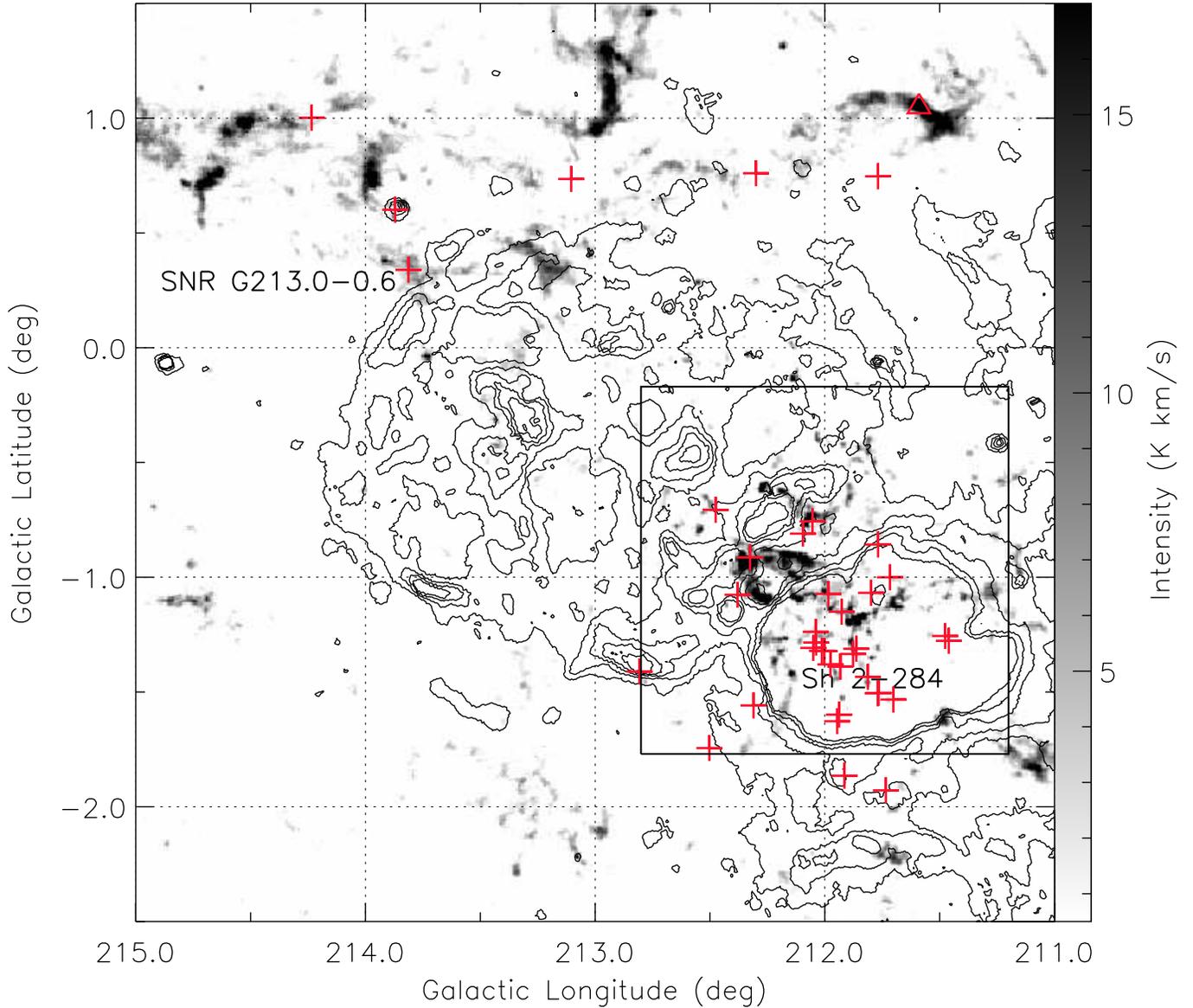}
\caption{
Integrated \twCO\ ($J$=1--0) emission toward SNR G213.0$-$0.6 from 
35 to 54~km~s$^{-1}$, overlaid with the same contours as in Figure 
\ref{G213rgb}. The red crosses indicate positions of early-type OB stars, 
which are probably associated with these MCs. The red triangle shows the
maser source G211.59$+$01.05 at a parallax distance of 4.4 kpc 
\citep{2014ApJ...783..130R}. The lower-right box indicates the region 
shown in Figure \ref{G213sub}.
\label{G213_OBsub}}
\end{figure*}

\begin{figure*}
\includegraphics[trim=-10mm 0mm 0mm 100mm,scale=0.8,angle=0]{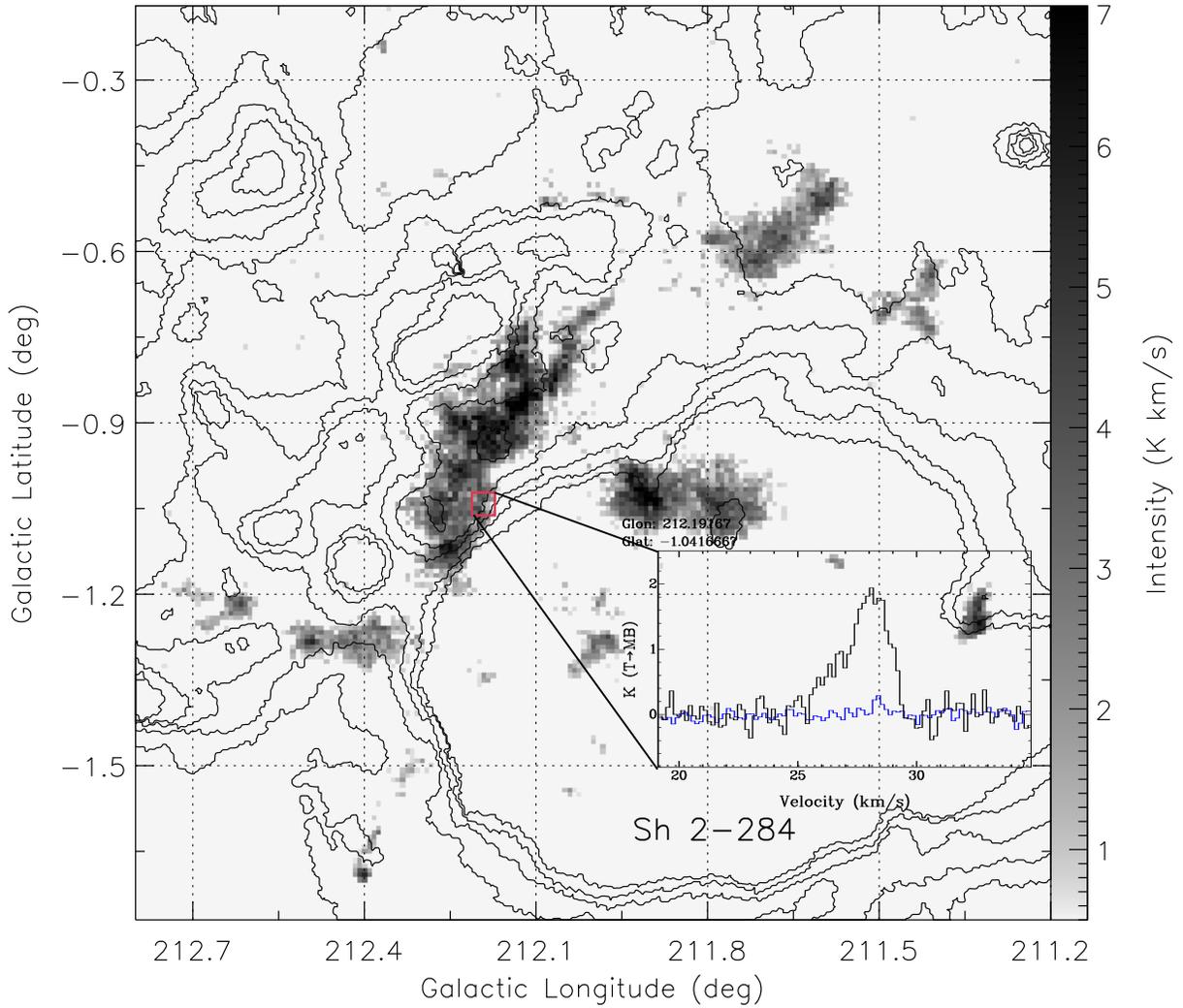}
\caption{
Integrated \twCO\ ($J$=1--0) emission toward the southwestern boundary of
SNR G213.0$-$0.6 from 24 to 33~km~s$^{-1}$,
overlaid with the same contours as in Figure \ref{G213rgb}. The spectra
of position $l=$212\fdg192 and $b=-$1\fdg042 are also given (\twCO\ in black
and \thCO\ in blue).
\label{G213sub}}
\end{figure*}

On the other hand,
MCs at $V_{\rm LSR}$=42--48$\km\ps$ (in the velocity interval of 
35--54$\km\ps$, see Figure \ref{G213_OBsub}) are probably from
the Milky Way's Perseus arm and/or the Norma (Outer) arm at a 
kinematic distance of $\sim$4.5--5.5 kpc 
\cite[e.g., refer to the rotation model of][]{2014ApJ...783..130R}. 
These MCs are clearly associated with \HII\ region Sh 2-284 
and a group of early-type OB stars (see the region of Sh 2-284 in this figure).
We also find that molecular gas at $V_{\rm LSR}$=28$\km\ps$ 
(in the velocity interval of 24--33$\km\ps$, see Figure \ref{G213sub}) 
seems to be located between the southwestern boundary of
SNR G213.0$-$0.6 and \HII\ region Sh 2-284. The 28$\km\ps$ MC 
component is probably also related to the \HII\ region Sh 2-284
because the blueshifted CO gas with respect to the 42--48$\km\ps$ 
molecular component has corresponding dark patches against the 
bright optical emission. The molecular gas presents blueshifted 
wing profiles on the side of the cloud facing \HII\ region Sh 2-284
(e.g., see the red box and its spectra in Figure \ref{G213sub}).
Accounting for the low metallicity in the \HII\ region, the distance 
of Sh 2-284 is estimated to be about 4 kpc
\citep{2010A&A...509A.104D,2011MNRAS.410..227C} 
or 4.5 kpc \citep{2015A&A...584A..77N}, which is in good 
agreement with the kinematic distance from our CO observations 
(e.g., the MCs with $V_{\rm LSR}\sim$42--48$\km\ps$ physically 
associated with the \HII\ region Sh 2-284). 
However, we do not find any convincing evidence of SNR--MC 
interaction for such molecular components.

Finally, many MCs outside the SNR can be seen in
the north of Figure \ref{G213_OBsub}, 
in which several early-type OB stars seem to be related to 
these MCs. For example, B0 star HD 289291 (211\fdg768,$+$0\fdg747,
the upper-right cross in Figure \ref{G213_OBsub}) is very likely 
responsible for its northern MC shell at $V_{\rm LSR}\sim$45$\km\ps$ 
(see the upper-right corner in Figure \ref{G213_OBsub}). 
It is interesting to note that a maser source G211.59$+$01.05
at $V_{\rm LSR}\sim45\pm5\km\ps$ 
(see red triangle in Figure \ref{G213_OBsub})
is exactly located in the MC shell.
The parallax distance of the maser is about 4.4 kpc 
\citep[0.228$\pm$0.007 mas in Table 1 of][]{2014ApJ...783..130R}. 
Due to the close correspondence between the molecular gas and the 
maser source, we suggest that the northern and southwestern MCs in
the interval of $V_{\rm LSR}\sim$42--48$\km\ps$ (and their associated 
OB stars, see red crosses in Figure \ref{G213_OBsub}) are at the 
same distance of 4--5 kpc. Accordingly, SNR G213.0$-$0.6 has no 
connection to \HII\ region Sh 2-284 (and the $V_{\rm LSR}\sim$28$\km\ps$ 
MCs) because of their different distances.

\section{SUMMARY}
A large-scale CO survey has been performed with high spatial 
($\sim50''$) and velocity ($\sim0.2\km\ps$) resolution to study the 
molecular environment of SNRs G205.5$+$0.5, G206.9$+$2.3, and 
G213.0$-$0.6. The high-quality CO data reveal complex structures 
in the molecular gas toward the three remnants. The resulting CO 
maps also allow us to investigate in detail the relationship between 
the SNRs and the molecular gas. We summarize our main results as follows.

1. SNR G205.5$+$0.5 is actually associated with the MCs at 
$V_{\rm LSR}\sim$5$\km\ps$ and 19$\km\ps$. The 5$\km\ps$ 
molecular gas, which exhibits the blueshifted CO broadening, 
is identified to be connected with obscuring optical regions. 
Whereas the 19$\km\ps$ MCs with redshifted line-broadening
are on the farside of the remnant. On the other hand, 
the nearby Rosette Nebula is also physically related to these
two MC components, suggesting the association between the
Nebula and the SNR.
Using the 3D extinction map from \cite{2015ApJ...810...25G}, 
the distances of the two molecular components are estimated
to be $\sim$1.6 kpc for the 5$\km\ps$ MCs and $\sim$1.6--2.0 kpc for the
19$\km\ps$ MCs, which agree well with the distance of $\sim$1.6~kpc 
from the connection between the SNR and the Rosette Nebula.
The confirmation of SNR--MC 
interaction toward SNR G205.5+0.5 supports the hadronic 
origin of the extended gamma-ray emission inside the remnant.

2. SNR G213.0$-$0.6 is interacting with the 8--11$\km\ps$ MCs. 
Based on the 3D extinction map from \cite{2015ApJ...810...25G},
the SNR's distance is $\sim$1~kpc from that of the associated molecular 
gas, which is consistent well with the estimates from the updated 
$\Sigma$--$D$ relationship \citep{2014SerAJ.189...25P} and the kinematic 
distance inferred from the LSR velocity of 8--11$\km\ps$.
The remnant has no relation to the southwestern \HII\ 
region Sh 2-284, which is located at a distance of 4--5 kpc.

3. For SNR G206.9$+$2.3, we notice that this remnant seems to be 
located in a molecular cavity at $\sim$15$\km\ps$.
Moreover, a partial shell structure at $V_{\rm LSR}\sim$14--15$\km\ps$ 
is found to be expanding outward at the southeastern boundary of the 
remnant. While this evidence may indicate an association between the SNR 
and the molecular gas, no significant molecular line broadening has 
been obtained. If the SNR--MC association is true, the kinematic distance 
to SNR G206.9$+$2.3 is $\sim$1.6~kpc.

4. SNRs G205.5$+$0.5 and G213.0$-$0.6 evolved in the low-density 
environment and are now interacting with their ambient molecular gas.
However, SNR G206.9$+$2.3 seems to be still evolving in the molecular
cavity. The SNRs' low-density environment may be created by their 
progenitors' activities. These SNRs are a good laboratory 
for establishing the relationships between SNRs and their environments
because of their large angular sizes, relatively near distances, and 
likely SNR--MC interaction. Detailed analyses and discussions will be
valuable for understanding the remnant's evolution and the interaction 
of SNRs with their surroundings through further multiwavelength studies.

\acknowledgments
The authors acknowledge the staff members of the Qinghai Radio
Observing Station at Delingha for their support of the observations.
We would like to thank the anonymous referee for valuable comments
and suggestions that helped to improve this paper. 
This work is supported by NSFC grants 11233001 and 11233007. 
X.Z. acknowledges support by NSFC grant 11403104 and Jiangsu Provincial Natural
Science Foundation grant BK20141044. Y.C. acknowledges support by
973 Program grant 2015CB857100 and NSFC grant 11633007.
Y.L. acknowledges support by NSFC grant 11233006.

\bibliographystyle{aasjournal}
\bibliography{references}

\clearpage

\clearpage

\clearpage

\clearpage

\clearpage

\clearpage

\clearpage

\clearpage

\clearpage

\clearpage

\clearpage

\end{document}